\documentclass[1p]{elsarticle}

\usepackage{hyperref,graphicx,array,url,color}
\usepackage{multirow}
\usepackage{graphicx}
\usepackage{algorithm}
\usepackage[noend]{algpseudocode}
\usepackage{subfigure}
\usepackage{caption}
\usepackage{amsmath,bm}

\begin{document}

\begin{frontmatter}

\title{\textbf{Learning Shared Semantic Space with Correlation Alignment
 for Cross-modal Event Retrieval}}

\author[rvt,bb]{Zhenguo Yang\corref{cor1}}
\ead{zhengyang5-c@my.cityu.edu.hk}
\author[rvt]{Zehang Lin}
\ead{gdutlin@outlook.com}
\author[rvt]{Peipei Kang}
\ead{ppkanggdut@126.com}
\author[cc]{Jianming Lv}
\ead{jmlv@scut.edu.cn}
\author[dd]{Qing Li}
\ead{csqli@comp.polyu.edu.hk}
\author[rvt]{Wenyin Liu\corref{cor1}}
\ead{liuwy@gdut.edu.cn}

\cortext[cor1]{Corresponding authors}

\address[rvt]{Department of Computer Science and Technology, Guangdong University of Technology, Guangzhou, China}
\address[bb]{Department of Computer Science, City University of Hong Kong, Hong Kong, China}
\address[cc]{School of Computer Science and Engineering, South China University of Technology, Guangzhou, China}
\address[dd]{Department of Computing, The Hong Kong Polytechnic University, Hong Kong, China}

\begin{abstract}
 In this paper, we propose to learn shared semantic space with correlation alignment (${S}^{3}CA$) for multimodal data representations, which aligns nonlinear correlations of multimodal data distributions in deep neural networks designed for heterogeneous data. In the context of cross-modal (event) retrieval, we design a neural network with convolutional layers and fully-connected layers to extract features for images, including images on Flickr-like social media. Simultaneously, we exploit a fully-connected neural network to extract semantic features for texts, including news articles from news media. In particular, nonlinear correlations of layer activations in the two neural networks are aligned with correlation alignment during the joint training of the networks. Furthermore, we project the multimodal data into a shared semantic space for cross-modal (event) retrieval, where the distances between heterogeneous data samples can be measured directly. In addition, we contribute a Wiki-Flickr Event dataset, where the multimodal data samples are not describing each other in pairs like the existing paired datasets, but all of them are describing semantic events. Extensive experiments conducted on both paired and unpaired datasets manifest the effectiveness of ${S}^{3}CA$, outperforming the state-of-the-art methods.
\end{abstract}


\end{frontmatter}

\section{Introduction}
News events (e.g., festivals, politics, natural disasters, etc.) are always happening around the world, which can be recorded and reported in social media platforms (e.g., Flickr, Twitter, etc.) and news media sites (e.g., BBC News, Yahoo News, etc.) in multimodal data forms, such as texts, images and videos, etc. In turn the multimodal data on the Internet platforms are not independent with each other, and there are supposed to be certain relationships among them. For instance, we may browse an image shared by a user on Flickr depicting an influential real-world event. It is possible that we can find news reports from news media that are related to the same event. The data distributed on different platforms can be related to the same event happenings. In reality, when we know a news event from a platform, e.g., social media, we may be also curious about the related content about the event in other platforms, e.g., news media. On one hand, social media data shared by amateur users provide personal and non-official viewpoints of events in free forms, while news media contributed by professional journalists are from more official perspectives in formal forms. On the other hand, different data modalities can be excellent at expressing and conveying specific aspects of facts and events. These two aspects are the motivation of the scenario of Cross-modal Event Retrieval \cite{1}.

The scenario of cross-modal event retrieval can be illustrated as that, given a data query describing an event, it aims to retrieve the data samples related to the same event from a database. It differs in two aspects compared to the traditional cross-modal retrieval. On one hand, the query and the data samples in the database can be in different data modalities from different data domains, e.g., a textual article from news media as a query, and images from social media constitute the database for retrieval. On the other hand, for the strongly-aligned paired data in Fig. 1-a and Fig. 1-b, the textual contents are exact descriptions of the visual images in ``injective mapping". For the weakly-aligned unpaired data in Fig. 1-c, the texts and images are contributed by different social media users and journalists, while they are describing the same high-level event concepts, which can be seemed as ``many-to-many associations". To the best of our knowledge, it is the seminal work investigating cross-modal retrieval tasks on weakly-aligned unpaired data, especially focusing on real-world events.

\begin{center}
\includegraphics[width=370pt]{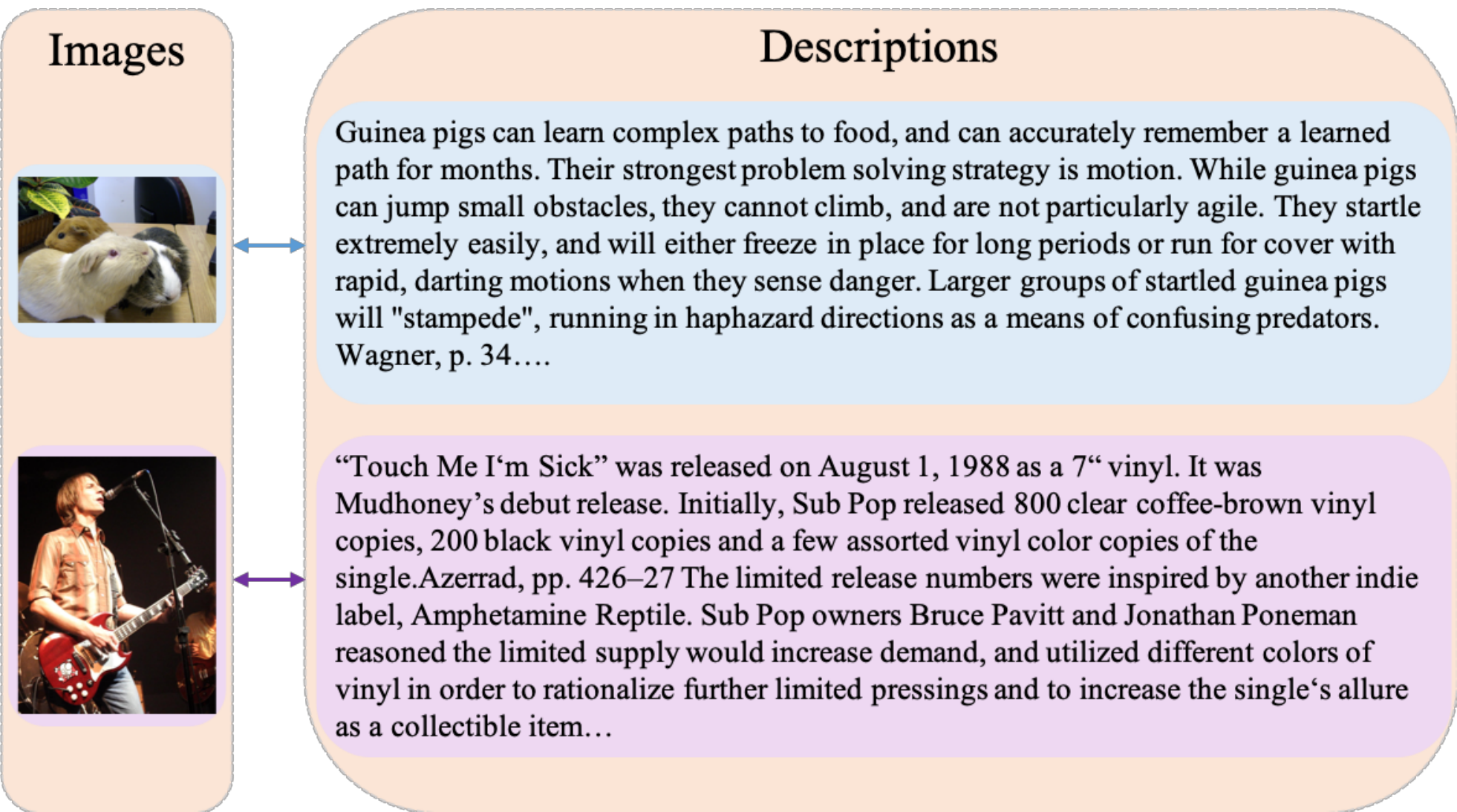}
\label{fig:f1-a}
\end{center}

\begin{center}
\textbf{(a)} (Strongly-aligned paired) Wikipedia dataset
\end{center}

\begin{center}
\includegraphics[width=370pt]{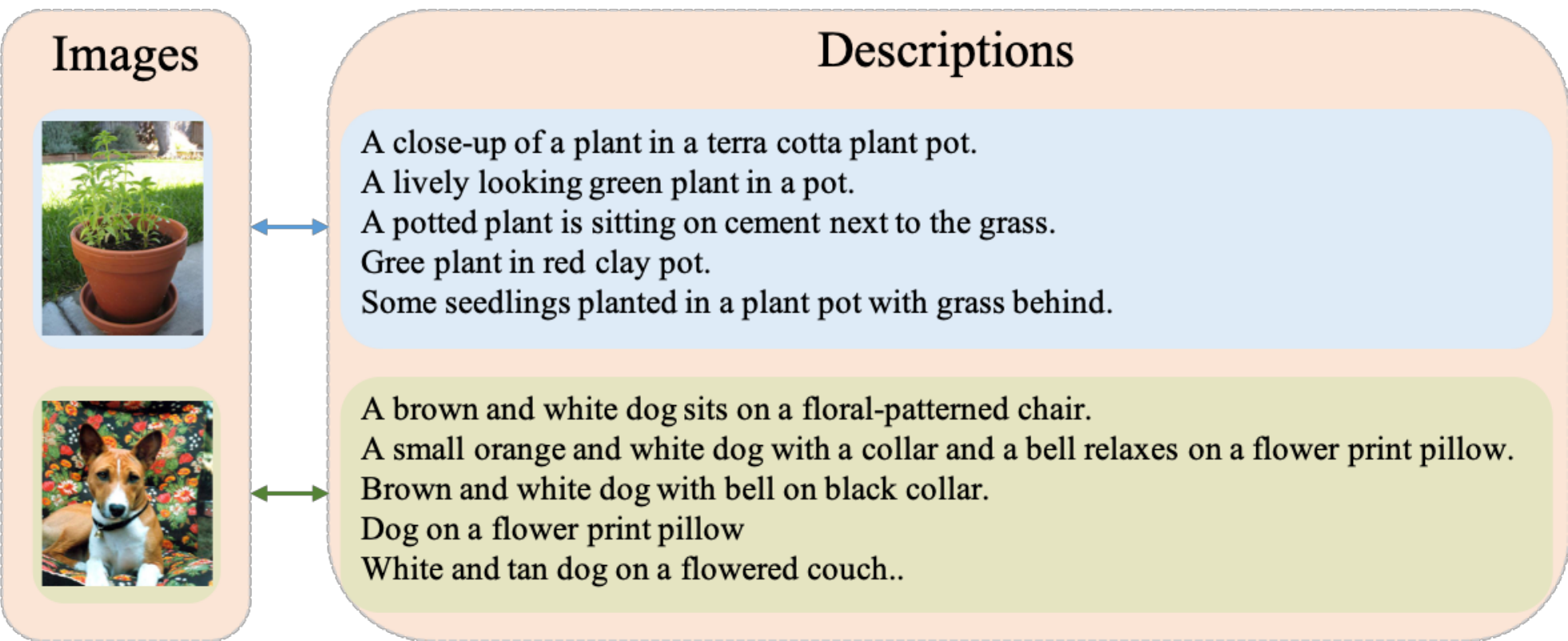}
\label{fig:f1-b}
\end{center}

\begin{center}
\textbf{(b)} (Strongly-aligned paired) Pascal Sentence dataset
\end{center}

\begin{center}
\includegraphics[width=370pt]{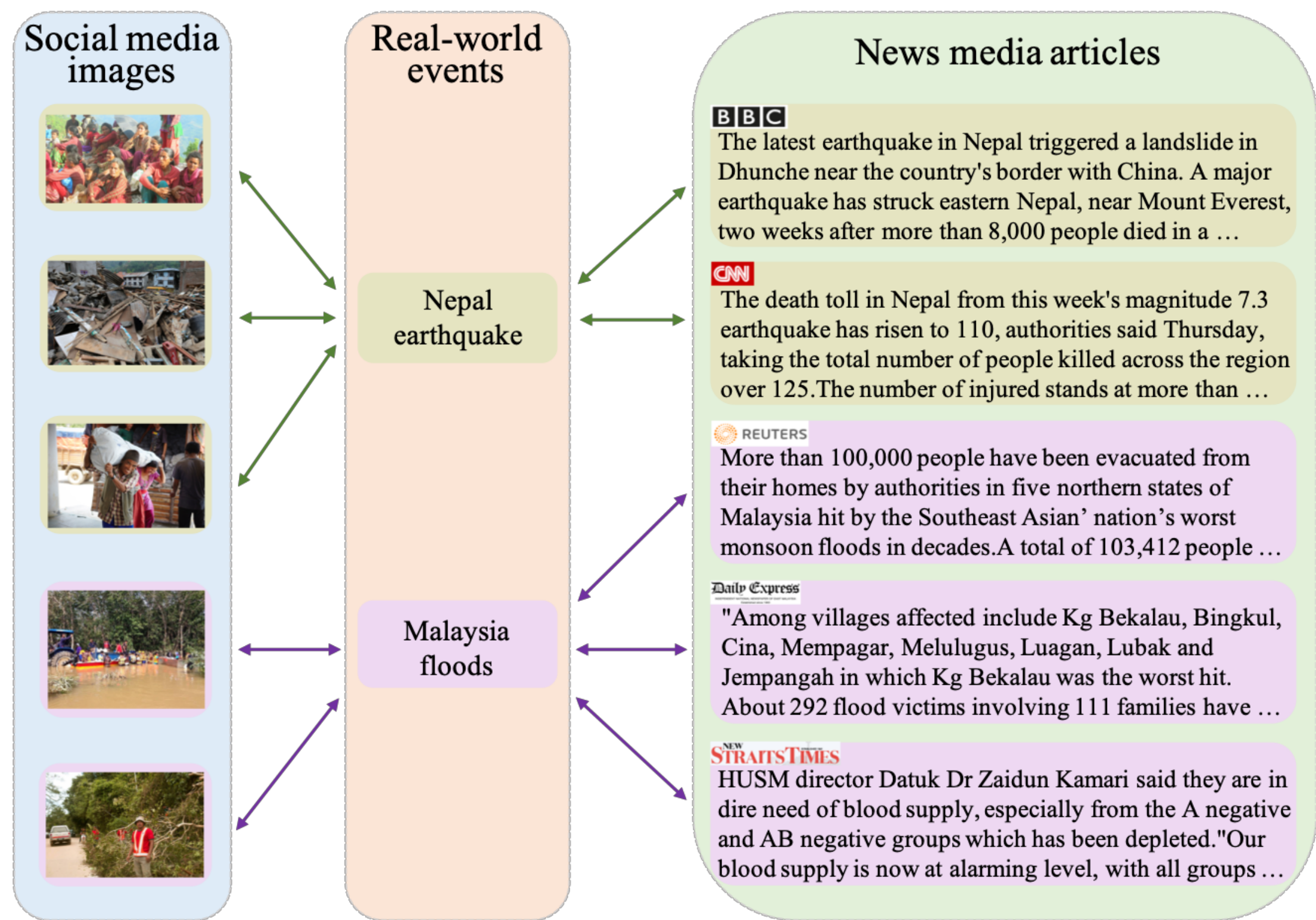}
\label{fig:f1-c}
\end{center}

\begin{center}
\textbf{(c)} (Weakly-aligned unpaired) Wiki-Flickr Event dataset
\end{center}

\noindent \textbf{Fig. 1.} Examples of strongly-aligned image-text pairs from Wikipedia dataset, Pascal Sentence dataset, and weakly-aligned unpaired examples from our Wiki-Flickr Event dataset. The corresponding text is the exact description of an image in the strongly-aligned data pairs. In contrast, the weakly-aligned textual content does not describe an image exactly, but they share the same event label. Note that the images are collected from social media, while the news articles are collected from different news media in subfigure (c).

In the context of cross-modal event retrieval, a few challenging issues need to deal with. Firstly, the heterogeneity of the data makes it impossible for machines to measure the difference between the multimodal data directly. Taking texts and images as an example, images are the representation of content (e.g., events, and objects) in visual perception in pixels forms, while texts are human's descriptions at the language level in words forms. Pixels and words are different both in data dimensions and distributions, which are not comparable. However, they have certain underlying relations intuitively if they describe the same events or objects. Modeling the intrinsic structures underlying the heterogeneous data is highly expected. Secondly, the unpaired data are harder to be associated with each other compared with paired data, because the data samples are not exact descriptions between each other. On one hand, modality-independent learning models may lose the relations among the cross-modal data samples. On the other hand, paired-data-based training strategies may be distracted by the low-quality data pairs in unpaired datasets. Joint training mechanisms for learning multimodal data representation considering the characteristics of unpaired data are highly expected.

To address the aforementioned challenges, numerous methods have been proposed in the context of cross-model retrieval, which can be divided into three categories: statistical correlation models, ranking models, and deep neural networks (DNN) based models. 1). Statistical correlation models are designed to align the cross-modal data with linear projection matrices, and learn common spaces where the correlations between modalities are maximized; 2). Ranking models design objectives to make the retrieved relevant pairs get higher rankings compared to the irrelevant pairs; 3). DNN-based models take the advantage of DNN for feature extraction or feature generation, and usually introduce some constraints on hidden units of the multimodal DNNs during the training processes. Overall, deep learning models have achieved significant improvements on the performance, showing a thriving trend in cross-model retrieval. However, the aforementioned models focus on paired datasets \cite{2, 3, 4}, which have limitations of expressing complicated semantic concepts like events. For an image (or a sentence), it only describes specific aspects of events partially, such as when, where, who, how, etc. The existing cross-modal retrieval methods have not taken into account the unpaired data.

This paper extends our previous work \cite{1} by proposing to learn shared semantic space with correlation alignment for multimodal data fusion, denoted as ${S}^{3}CA$. As an extension, ${S}^{3}CA$ improves the previous work mainly in three aspects. 1). Instead of learning modality-specific data representations independently, ${S}^{3}CA$ achieves a shared semantic space by training the modality-specific neural networks jointly in an interactive manner. 2). Numerous interactive regularization terms have been introduced and investigated during the joint training of the neural networks, which align the distributions of multi-domain and multimodal data with nonlinear transformations.  ${\rm 3).\ More}$ public datasets in addition to our Wiki-Flickr Event dataset have been investigated to illustrate the effectiveness of the proposed ${S}^{3}CA$ and numerous regularization terms. The Wiki-Flickr Event dataset has been released for public use on GitHub (https://github.com/zhengyang5/Wiki-Flickr-Event-Dataset).

The main contributions of this paper are summarized as follows.
\begin{enumerate}[    (1)]
\item We advocate the problem of cross-modal event retrieval on weakly-aligned unpaired data, breaking through the limitations of cross-modal retrieval focusing on strongly-aligned data pairs alone.
\item We propose to learn shared semantic space for heterogeneous data with correlation alignment, which aligns nonlinear correlations of layer activations in modality-specific neural networks in an interactive joint training manner.
\item We investigate numerous interactive regularization terms on multimodal data alignment that can be adopted by modality-specific neural networks, which manifests the effectiveness of correlation alignment in the context of cross-modal (event) retrieval.
\item We have collected a real-world dataset for research on cross-modal (event) retrieval, consisting of both images shared by amateur social media users and news articles contributed by journalists from various news media sites. The dataset has been released, which can hopefully be used to promote the research on this topic and advance related applications.
\end{enumerate}

The rest of this paper is organized as follows. Section 2 reviews the related works on cross-modal retrieval. Section 3 introduces the preliminaries. Section 4 presents the proposed ${S}^{3}CA$, followed by the experiments and analyses in Section 5. Finally, Section 6 offers some concluding remarks.
\section{Related Work}
We investigate a number of widely-used and recent cross-modal retrieval models from the categories of statistical correlation models, ranking models, and DNN-based models, respectively.

\subsection{Statistical correlation models}

Statistical correlation models are designed to learn a subspace where cross-modal data are aligned from the perspective of statistics. Canonical correlation analysis (CCA) \cite{5} is a representative work, which projects two sets of data to a common subspace where their correlations are maximized. Similar to CCA, cross-modal factor analysis (CFA) \cite{6} minimized the Frobenius norm between the transformed cross-modal data. Kernel-CCA \cite{5} introduced kernel functions for nonlinear correlations, which is the kernel extension of CCA. Rasiwasia et al. \cite{7} learned a semantic space using CCA representation and semantic category information for cross-modal retrieval tasks. Sharma et al. \cite{8} proposed a generalized multi-view analysis (GMA) as a supervised extension of CCA. Multi-view CCA \cite{9} extended CCA by incorporating the high-level image semantic keywords as the third view. Multi-label CCA \cite{10} took the multi-label annotations to establish correspondences, without relying on the pairwise modalities like CCA. As pointed out by Tran et al. \cite{11}, using CCA directly may lead to coarse subspace, and the relationships between real data are too complicated to be captured by linear projections alone.

\subsection{Ranking models}

The intuition of ranking models is that relevant pairs in the retrieved results should rank higher than the irrelevant pairs \cite{12}. Recently, neural networks combined with ranking loss become popular in the context of cross-modal retrieval. Wang et al. \cite{13, 14} designed a two-branch neural network with multiple layers of linear projections followed by nonlinearities, and learned the joint embeddings for images and texts with a large margin objective which combines ranking constraints and neighborhood structure preserving constraints. Salvador et al. \cite{15} introduced a large-scale recipe dataset, and jointly learned the embeddings of images and recipes in a common space by maximizing the cosine similarity between positive recipe-image pairs and minimizing the similarity between negative pairs. Zhang et al. \cite{16} designed a sampling strategy and define discriminative ranking loss on two heterogeneous networks to obtain discriminative embeddings for cross-modal retrieval. However, the training of ranking loss relies on high-quality and unambiguous data pairs, which may not be appropriate for unpaired datasets, like our Wiki-Flickr Event dataset.

\subsection{DNN-based models}
DNN-based models exploit neural networks to extract nonlinear features, which can also combine with the previous strategies, such as deep-CCA \cite{17} maximized the correlation by deep neural networks. Feng et al. \cite{18} proposed the correspondence autoencoder (Corr-AE) by conducting two uni-modal autoencoders by learning representation and correlation. Peng et al. \cite{19} proposed the cross-media multiple deep network (CMDN) to learn cross-modal shared representation by a hierarchical architecture with networks. Cross-modal correlation learning (CCL) \cite{20} learned modality-specific representation firstly, and then leveraged the intra-modality semantic constraint and inter-modality pairwise constraint. He et al. \cite{21} adopted two convolutional networks to learn a common space, where the likelihoods of all matched pairs were maximized in an end-to-end manner. Zhang et al. \cite{16} fine-tuned the ResNet and LSTM networks to encode images and texts respectively, and optimized the network with a discriminant ranking loss. Wei et al. \cite{3} proposed a deep semantic matching method (deep-SM) to transform the problem of common space learning to classification. Fan et al. \cite{22} used LSTM to generate language descriptions of images, and mapped images and texts to a semantic space. Wang et al. \cite{23} used adversarial learning to produce modality-invariant and discriminative representations. SSAH \cite{4} used two adversarial networks to force the modality-specific features consistent with the semantic features. Zhang et al. \cite{24} introduced the attention mechanism and adopt GANs to generate the attention distributions and learn the binary codes. The DNN-based models are becoming the mainstream in the context of cross-modal retrieval, benefiting from the superior ability of neural networks for feature learning, which motivates us to explore the deep models for cross-modal event retrieval.

\section{Preliminaries and Problem Statement}

Given a set of $n$ samples in an image database $D_I$ and $m$ samples in a text database $D_T$, which are related to a number of $K$ real-world events, the images and texts can be collected from different data domains, such as images from Flickr-like social media, and textual articles from news media (e.g., BBC News, Yahoo News, etc.). In particular, the images and texts can be unpaired, i.e., they are not trying to describe each other but describing high-level semantic concepts like events jointly. The raw features of an image \emph{I} and a text \emph{T} can be denoted as \emph{${R}_{I}$} and \emph{${R}_{T}$}, respectively. The main notations are described in Table 1. A more formal definition of the current problem is illustrated as follows.

\textbf{Problem 1 (Cross-modal Event Retrieval).} Cross-modal event retrieval aims to obtain a latent space by learning transformations $\varphi$ and $\Psi$ for images and texts, which map \emph{${R}_{I}$}  and \emph{${R}_{T}$}  to semantic embeddings \emph{${S}_{I}$}  and \emph{${S}_{T}$} , i.e., $\mathcal{S_I}=\Psi(R_I)$, and $\mathcal{S_T}=\Psi(R_T)$, respectively. In the new space, taking \emph{${S}_{I}$} as a query to rank the text samples in \emph{${D}_{T}$} by using a predefined distance measurement $dist$ on the cross-model embeddings, we can obtain a ranking list of all the text samples. Ideally, the top \emph{${K}_{I}$} samples in the list are associated with the same event label with the query \emph{${S}_{I}$}, where \emph{${K}_{I}$} is the exact number of texts in \emph{${D}_{T}$} with the same event label as \emph{${S}_{I}$}. Similarly, the query can also be a text sample.

\begin{center}
\textbf{Table 1.} Main notations
\end{center}

\begin{tabular}{p{1.5cm}p{10.1cm}}
\hline
\textbf{Notation}&\textbf{Description}\\
\hline
\emph{$fc1$}&The first fully-connected layer after the neural network for image\\
\emph{$fc2$}&The second fully-connected layer after the neural network for image\\
\emph{${fc1}^{'}$}&The first fully-connected layer in the neural network for text\\
\emph{${fc2}^{'}$}&The second fully-connected layer in the neural network for text\\
\emph{${D}_{I}$}&An image database consisting of $n$ samples\\
\emph{${D}_{T}$}&A text database consisting of $m$ samples\\
\emph{${X}_{I}$}&An image training set\\
\emph{${X}_{T}$}&A text training set\\
\emph{I}&An image\\
\emph{T}&A text\\
\emph{${R}_{I}$}&Raw representation of image \emph{I}\\
\emph{${R}_{T}$}&Raw representation of text \emph{T}\\
\emph{K}&Number of labels (events)\\
\emph{${O}_{I}$}&Output of \emph{I} after \emph{$fc2$}\\
\emph{${O}_{T}$}&Output of \emph{T} after \emph{${fc2}^{'}$}\\
\emph{${S}_{I}$}&Semantic embedding of \emph{I}\\
\emph{${S}_{T}$}&Semantic embedding of \emph{T}\\

\hline
\end{tabular}

\section{Methodologies}
In this section, the framework for cross-modal (event) retrieval and the details of the proposed ${S}^{3}CA$ model are introduced, which uses the convolutional network and fully-connected network to learn the shared semantic space for social media images and news media articles. Fig. 2 illustrates the overview of the proposed framework.

\begin{center}
\includegraphics[width=\textwidth]{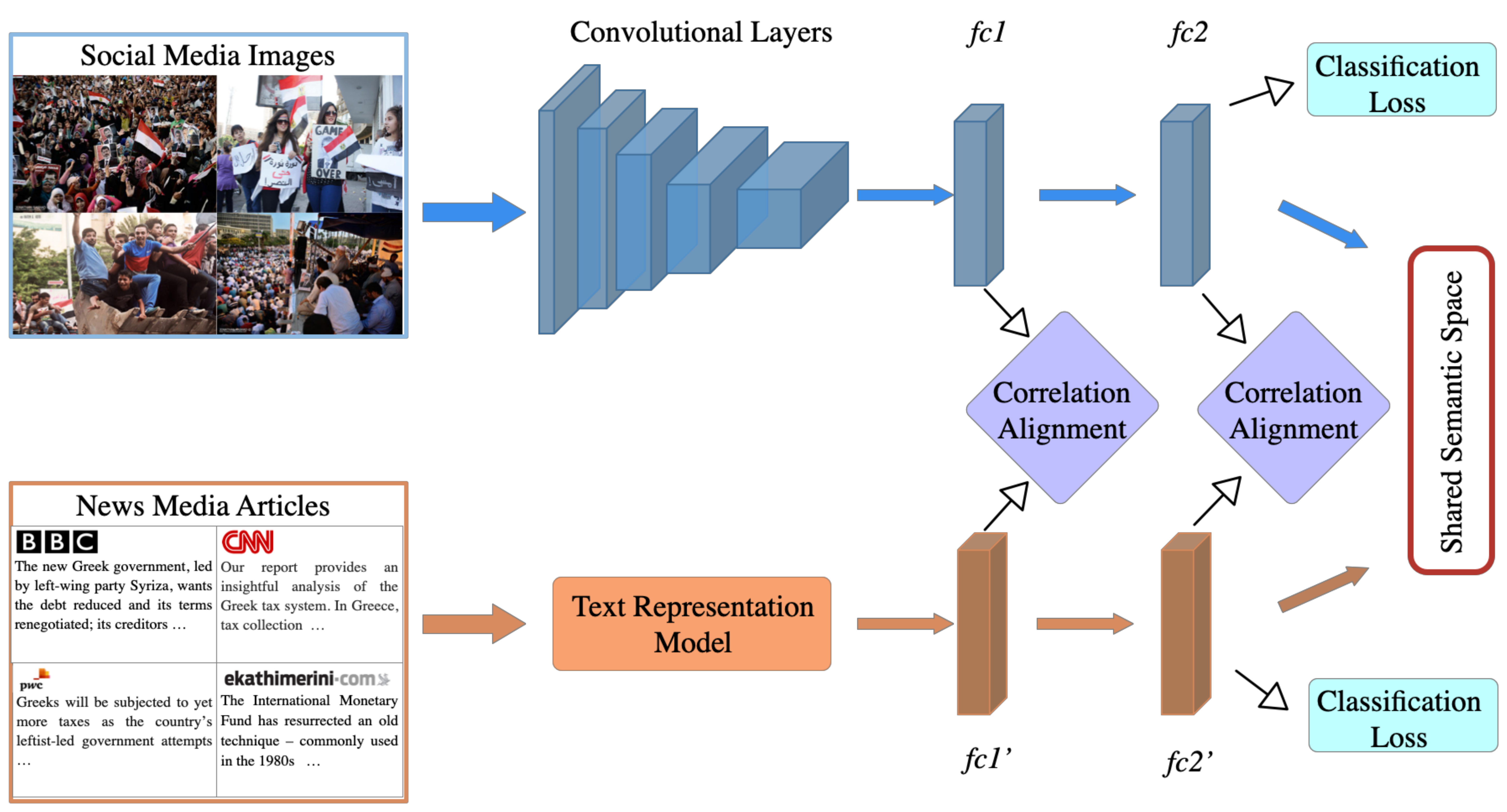}
\end{center}

\noindent \textbf{Fig. 2.} An overview of our proposed ${S}^{3}CA$. For images, we use a convolutional neural network to transfer semantic knowledge from ImageNet (in Section 4.1). For texts, we design a two-layer neural network to extract text semantics (in Section 4.2). We introduce correlation alignment to align the distributions of activation layers of the modality-specific neural networks (in Section 4.3). Finally, the multimodal data is embedded into shared semantic space for cross-modal retrieval (in Section 4.4).

\subsection{Learning Image Semantics with Knowledge Transfer}

\begin{center}
\includegraphics[width=\textwidth]{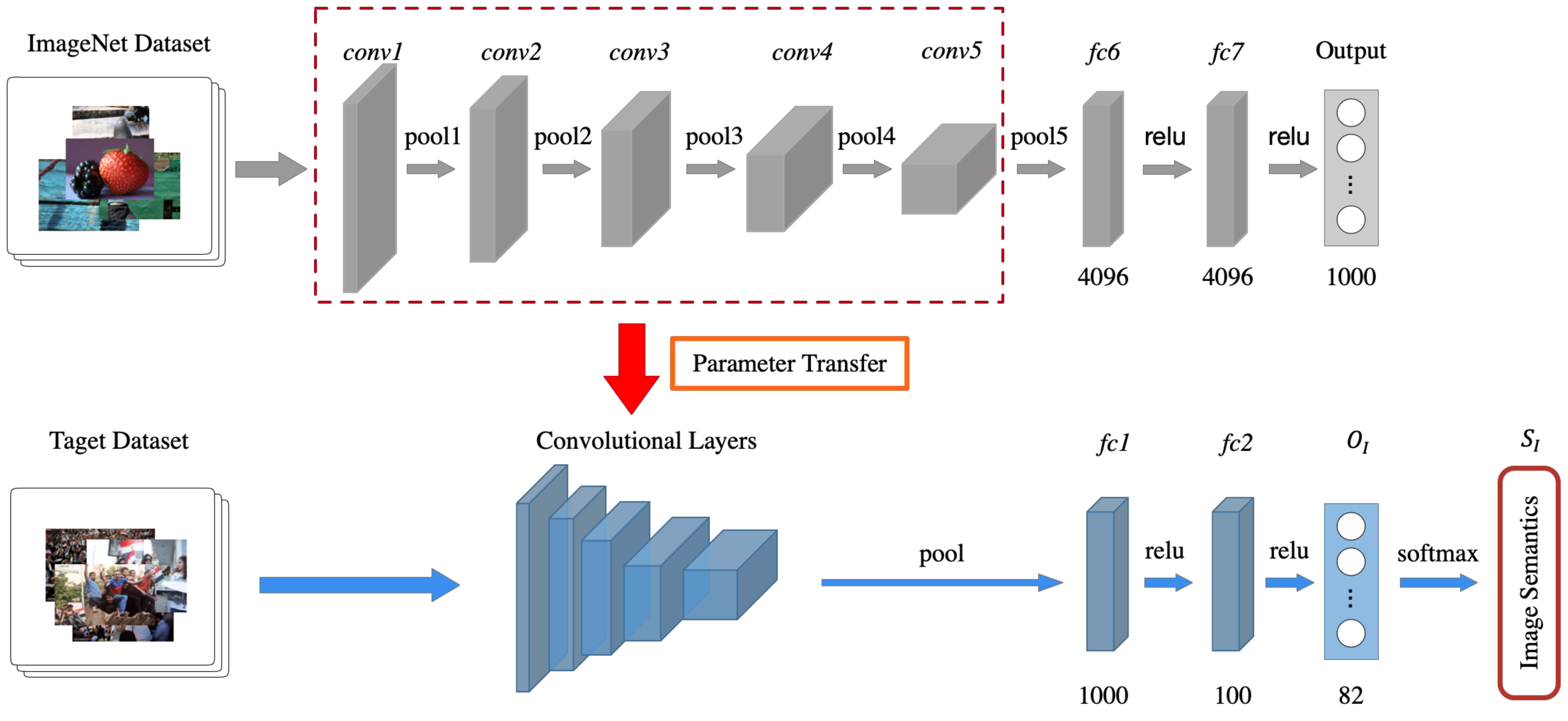}
\end{center}

\noindent \textbf{Fig. 3.} Learning image semantics with knowledge transfer.
\\

Inspired by the outstanding performance of convolutional neural network (CNN) on various recognition tasks, we propose to extract image semantics based on a pretrained VGG network \cite{25} on ImageNet. More specifically, we fine-tune the pre-trained CNN model on our target datasets to extract task-related visual features for the images. As shown in Fig. 3, we firstly replace the last two fully-connected layers in VGG, i.e., $fc6$ and $fc7$, with two randomly initialized fully-connected layers, i.e., \emph{$fc1$} and \emph{$fc2$}. Furthermore, we set the number of hidden units in the last fully-connected layer \emph{${O}_{I}$} to be the same with the number of categories in the target dataset. Take our Wiki-Flickr Event dataset as an example, there are 82 real-world events in total (refer to Section 5.1). Finally, a softmax function is exploited to obtain semantic embedding ${R}^K$ for image \emph{I}, where \emph{K} is the number of classes.
In terms of computations, the softmax function maps a $K$-dimensional vector $z$ to a $K$-dimensional vector $\sigma$(\emph{z}) of real values in the range of $(0, 1)$ that add up to 1. The image semantic embedding \emph{${S}_{I}$} is defined below:\

\begin{equation}
\label{eqn:L1}
({S}_{I})_j = P(y=j|I) = \frac{e^{{o_I}_j}}{\sum_{i=1}^{K}e^{{o_I}_i}}, \quad for \quad  j =1,...,K.
\end{equation}

\noindent where \emph{$P(y = j|I)$} represents the predicted probability for the \emph{j}-th class given an image sample \emph{I}. \emph{${S}_{I}$} $\in$ \emph{${R}^{K}$}  is the embedding vector of image semantic. (\emph{${S_I}_j$} represents the \emph{j}-th element in the vector.
In terms of the loss terms of the CNN model for the current tasks, we utilize the cross-entropy loss to optimize its parameters. Therefore, the cost function for image semantic learning is defined as follows:

\begin{equation}
\label{eqn:L2}
{Loss}_{I} = -\sum_{j=1}^{K}1(y=j)log({S_I}_j),
\end{equation}

\noindent where $y$ is the label, and 1$(y=j)$ is an indicator, which is equal to 1 if \emph{y} = \emph{j}, otherwise it is 0.

\subsection{Learning Text Semantics by Fully-connected Neural Network}

\begin{center}
\includegraphics[width=\textwidth]{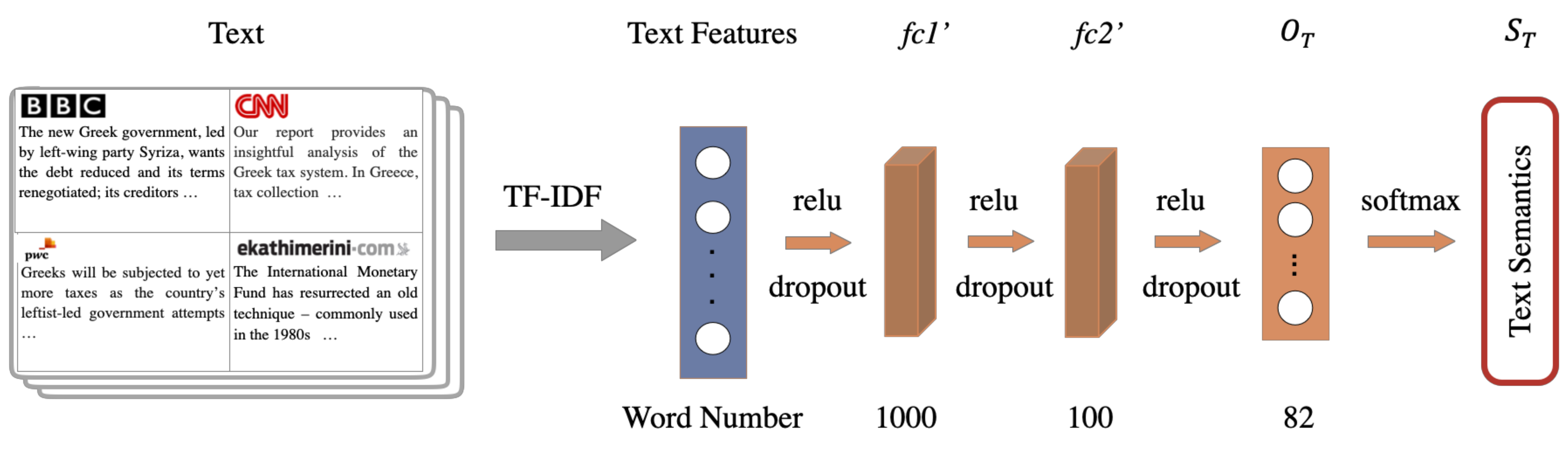}
\end{center}

\noindent \textbf{Fig. 4.} Learning text semantics by fully-connected neural network.
\\

For text, we apply term frequency-inverse document frequency (TF-IDF) to construct the textual representations. As shown in Fig. 4, we firstly obtain the raw text features by TF-IDF, before which stop words have been removed. The dimension of the vectors is equal to the number of tokens in the corpus. Furthermore, we utilize a 2-layer fully-connected network, i.e., \emph{${fc1}^{'}$} and \emph{${fc2}^{'}$}, to learn the hidden semantics underlying the documents, which is defined below:

\begin{equation}
\label{eqn:L3}
f(x) = max(0,x),
\end{equation}

\begin{equation}
\label{eqn:L4}
{h_t}^{(2)} = f^{(2)}({W_t}^{(1)} * R_T + {b_t}^{(1)}),
\end{equation}

\begin{equation}
\label{eqn:L5}
{o_T} = f^{(3)}({W_t}^{(2)} * {h_t}^{(2)} + {b_t}^{(2)}),
\end{equation}

\noindent where $f$($x$) represents the rectified linear unit (ReLU) function, i.e., the activation function, \emph{W} is the weight matrix, \emph{b} is the bias term, and \emph{${o}_{T}$} represents the output of the last fully-connected layer.
Finally, \emph{${o}_{T}$} is fed to a $K$-way softmax, which obtains semantic embedding \emph{${S}_{T}$} $\in$ \emph{${R}_{K}$} for a text \emph{T}. The text semantic embedding \emph{${S}_{T}$} is defined below:

\begin{equation}
\label{eqn:L6}
({S}_{T})_j = P(y=j|T) = \frac{e^{{o_T}_j}}{\sum_{i=1}^{K}e^{{o_T}_i}}, \quad for \quad  j =1,...,K.
\end{equation}

\noindent where \emph{$P(y = j|T)$} represents the predicted probability for the \emph{j}-th class given a data sample \emph{T}. \emph{${S}_{T}$} $\in$ \emph{${R}^{K}$}  is the text semantic embedding. \emph{${S_T}_j$} represents the \emph{j}-th element in the vector.

In terms of the loss terms of the fully-connected neural network to learn semantic embeddings for text, we also utilize the cross-entropy loss function in the following form:
\begin{equation}
\label{eqn:L7}
{Loss}_{T} = -\sum_{j=1}^{K}1(y=j)log({S_T}_j).
\end{equation}

\subsection{Deep Correlation Alignment}

1) \textbf{Correlation Alignment (CORAL)}. In our previous work, the modality-specific neural networks are trained independently, which may not take into account the inter-modality relationships. Therefore, we introduce an interactive regularization term in order to align the distributions between data representations achieved for texts and images, which is critical for cross-modal retrieval. We propose to minimize the difference in second-order statistics between the feature activations of the neural networks for texts and images by using CORAL \cite{26, 27}. As shown in Fig. 2, we introduce the CORAL constraints to align the distributions of the feature activations in the fully-connected layers of the neural networks, i.e., the output of \emph{${fc1}$} and \emph{${fc1}^{'}$}, and the output of \emph{${fc2}$} and \emph{${fc2}^{'}$}.

Without loss of generality, assume that \emph{${I}_{ij}$}  and \emph{${T}_{ij}$}  denote the \emph{j}-th dimension of the \emph{i}-th image and the \emph{j}-th dimension of the \emph{i}-th text, respectively. The CORAL loss can be defined as:

\begin{equation}
\label{eqn:L8}
{Loss}_{CORAL} = \frac{1}{4d^2}{\|C_I-C_T\|}^{2}_{F},
\end{equation}

\noindent where $d$ is the dimension of the input layer, \emph{${C}_{I}$} and \emph{${C}_{T}$} denote the feature covariance matrices, and {$\|.\|^{2}_{F}$} denotes the squared matrix of Frobenius norm. \emph{${C}_{I}$} and \emph{${C}_{T}$} can be obtained as follows:

\begin{equation}
\label{eqn:L9}
{C}_{I} = \frac{1}{n_I-1}{(I^TI-\frac{1}{n_I}I^T1^{n_I}I)},
\end{equation}

\begin{equation}
\label{eqn:L10}
{C}_{T} = \frac{1}{n_T-1}{(T^TT-\frac{1}{n_T}T^T1^{n_T}T)},
\end{equation}

\noindent where {$1^n$} denotes an {$n \times n$} matrix with all elements equal to 1.
The gradient for the input features can be calculated by the chain rule:

\begin{equation}
\label{eqn:L11}
\frac{\partial Loss_{CORAL}}{\partial I^{ij}} = \frac{1}{d^2(n_I-1)}{(I^T-\frac{1}{n_I}I^T1^{n_I})}^T{(C_I-C_T)}^{ij},
\end{equation}

\begin{equation}
\label{eqn:L12}
\frac{\partial Loss_{CORAL}}{\partial T^{ij}} = \frac{1}{d^2(n_T-1)}{(T^T-\frac{1}{n_T}T^T1^{n_T})}^T{(C_I-C_T)}^{ij}.
\end{equation}

2) \textbf{Discussions on relationships to existing methods}. In terms of modeling the interactions between modality-specific or domain-specific models, triplet loss \cite{28}, generative adversarial networks (GANs) \cite{23} and Maximum Mean Discrepancy (MMD) \cite{29} are thriving and representative interactive regularization terms that have been widely-used. More specifically, triplet loss relies on the qualities of the positive and negative data pairs in the training process, which aims to learn the rankings of the data pairs. Unlike triplet loss being limited to paired data, CORAL can be used to align multimodal data for both paired data and unpaired data.

In terms of the adversarial methods based on the idea of GANs \cite{30}, they usually introduce a modality classifier acting as ``Discriminator" to distinguish that the feature activations are from either the network trained on image domain or the network trained on the text domain. The adversarial methods aim to fool the Discriminators to make them unable to distinguish the aforementioned two cases, which indicate the feature activations are in the same data distribution. However, some recent works \cite{30, 31} reveal the high risk of failure that these methods are suffering from. Arora et al. \cite{32, 33} have pointed out that there are no theoretical guarantee that two domain-specific distributions of feature activations are becoming identical, even the discriminator is fully confused. In contrast, CORAL measures the difference between the feature activations in statistics directly.

In terms of MMD, it uses a polynomial kernel to transform images and texts (or data in source and target domains) into a common space, which can express arbitrary statistics of the data. On one hand, no previous work has proposed a closed form solution for MMD, while we can find the optimal solution for CORAL \cite{27}. On the other hand, the transformation of MMD is symmetric for both source and target domains, while CORAL transforms the feature activations for images and texts in an asymmetric manner, which is more flexible and often yields better performance on aligning the modality-specific distributions.

For comparisons in practice, we will investigate the performance of introducing the aforementioned interactive regularization terms in addition to CORAL in our experiments (refer to Section 5.5) for the cross-modal (event) retrieval tasks.

\subsection{Objective of ${S}^{3}CA$}
Consequently, combining the classification loss terms for semantic alignment in Sections 4.1 and 4.2 with the CORAL loss in Section 4.3, the objective function of ${S}^{3}CA$ is specified below:

\begin{equation}
\label{eqn:L13}
Loss = \frac{1}{m} \sum_{i=1}^{m} Loss_I+Loss_T+Loss_{CORAL}^{fc1 \sim fc1^{'}}+{Loss_{CORAL}^{fc2 \sim fc2^{'}}},
\end{equation}

\noindent where $m$ is the number of training samples, and the superscripts {$fc1 \sim fc1^{'}$} and {$fc2 \sim fc2^{'}$} denote CORAL constraint on the fully-connected layers between {$fc1$} and {$fc1^{'}$}, and the last fully-connected layers {$fc2$} and {$fc2^{'}$}, respectively.

To solve the objective function of ${S}^{3}CA$, stochastic gradient descent (SGD) can be used for optimizations as follows:

\begin{equation}
\label{eqn:L14}
{\theta}_I \gets {\theta}_I - \lambda \frac{\partial loss}{\partial {\theta}_I},
\end{equation}

\begin{equation}
\label{eqn:L15}
{\theta}_T \gets {\theta}_T - \lambda \frac{\partial loss}{\partial {\theta}_T},
\end{equation}

\noindent where {$\lambda$} represents the learning rate, {${\theta}_I$} and {${\theta}_T$} denote the parameters of the neural networks designed for images and texts, respectively.

For completeness, the training process of our proposed ${S}^{3}CA$ is summarized in Algorithm 1.

\begin{algorithm}[h]
\caption{The training process of ${S}^{3}CA$ Algorithm.}
\hspace*{0.02in} {\bf Input:}
image training data {$X_I$}; text training data {$X_T$}; parameters of neural network model for images {$\theta_I$}; parameters of neural network model for texts {$\theta_T$}; learning rate {$\lambda$}.\\
\hspace*{0.02in} {\bf Output:}
optimized ${S}^{3}CA$ model.
\begin{algorithmic}[1]

\State 	Pre-train {$\theta_I$} on ImageNet dataset
\State  {\bfseries repeat}
\State  \qquad Sample ${\{{{x_i}^I},{{y_i}^I}\}}_{i=1}^{m}$, ${\{{{x_i}^T},{{y_i}^T}\}}_{i=1}^{m}$ from {$X_I$}  and {$X_T$}
\State  \qquad Design an interactive deep model with objective in Equation (13)
\State  \qquad Compute stochastic gradient of {$\theta_I$} and {$\theta_T$} by following Equations (14) and (15)
\State \qquad Update Models:
\State \qquad ${\theta}_I \gets {\theta}_I - \lambda \frac{\partial loss}{\partial {\theta}_I}$
\State \qquad ${\theta}_T \gets {\theta}_T - \lambda \frac{\partial loss}{\partial {\theta}_T}$
\State {\bfseries until} ${S}^{3}CA$ converge\end{algorithmic}
\end{algorithm}

\subsection{Semantic Matching in the Shared Semantic Space}
Given the optimized ${S}^{3}CA$ model, we can obtain a shared semantic space for both images and texts by outputting the last fully-connected layers of the modality-specific neural networks, which take into account both intra-modality semantic information and inter-modality underlying relationships. For cross-modal (event) retrieval, distance metrics, e.g., Euclidean distance, cosine distance, Kullback-Leibler (KL) divergence, Normalized Correlation (NC), can be used to measure the distances between the instances of different modalities in the shared semantic space directly. In the experiments (refer to Section 5.6), we will investigate the influence of using various distance metrics in the context of cross-modal (event) retrieval.

\section{Experiments}
In this section, we evaluate the performance of our proposed approaches on both paired and unpaired datasets, and compare with several state-of-the-art algorithms.

\subsection{Dataset}
\begin{enumerate}[  1)]
\item \textbf{Wiki-Flickr Event dataset:} For cross-modal event retrieval, we collect 28,825 images from social media Flickr and 11,960 text articles from hundreds of news media, such as BBC News, The New York Times, Yahoo News, Google News, etc. The images and texts are not paired with each other, but they are related to 82 real-world events, such as ``2014 Hong Kong protests", ``Tianjin Explosion", ``Israeli legislative election, 2015", ``Shooting of Michael Brown", etc. Some examples are shown in Fig.1. The event labels cover a wide range of event categories (or event types) like emergency, natural disaster, sport, ceremony, election, protest, military intervention, economic crisis, etc.

We collect the dataset considering the principles \cite{Zhenguo2019} of high relevance in supporting the application needs, wide range of event types, non-ambiguity of the event labels, imbalance of the event clusters, and difficulty of discriminating the event labels, etc. For each event label, there is a corresponding Wikipedia entry. Some examples in our dataset are shown in Fig. 1, and the statistics of the dataset is shown in ${\rm Fig.\ 5}$. For data partitions, 60\% of the data samples are used as training set, 15\% of the data samples are used as validation set, and the rest 25\% are used as testing set. The dataset has been released to the public.

\item \textbf{Wikipedia dataset \cite{7}:} It contains 2,866 image-text pairs of 10 categories, which is widely used for cross-modal retrieval. By following \cite{18, 19}, we randomly split it into three parts: 2,173 pairs as training set, 231 pairs as validation set, and 462 pairs as testing set.

\item \textbf{Pascal Sentence dataset \cite{34}: }  It contains 1,000 images from 20 categories, and each image has 5 corresponding sentences as exact descriptions. For each category, we randomly select 40 image-description pairs as training set, 5 image-description pairs as validation set, and 5 image-description pairs as testing set by following \cite{18, 19}. Table 2 summarizes the statistics of the three datasets.
\end{enumerate}

\begin{center}
\textbf{Table 2.} Data partitions of the datasets.
\end{center}

\begin{tabular}{|l|l|l|l|l|l|l|l|}
\hline
\multirow{2}{*}{Dataset}&\multirow{2}{*}{\#labels}&\multicolumn{2}{|c|}{Training set}&\multicolumn{2}{|c|}{Validation set}&\multicolumn{2}{|c|}{Testing set}\\
\cline{3-8}
 & &Image&Text&Image&Text&Image&Text\\
\hline
Wikipedia&10&2,173&2,173&231&231&462&462\\
\hline
Pascal Sentence&20&800&4,000&100&500&100&500\\
\hline
Wiki-Flickr Event&82&17,295&7,176&4,324&1,794&7,206&2,990\\
\hline
\end{tabular}

\begin{figure}[h]
\centering
\subfigure[Social media images]{
\includegraphics[width=0.48\textwidth]{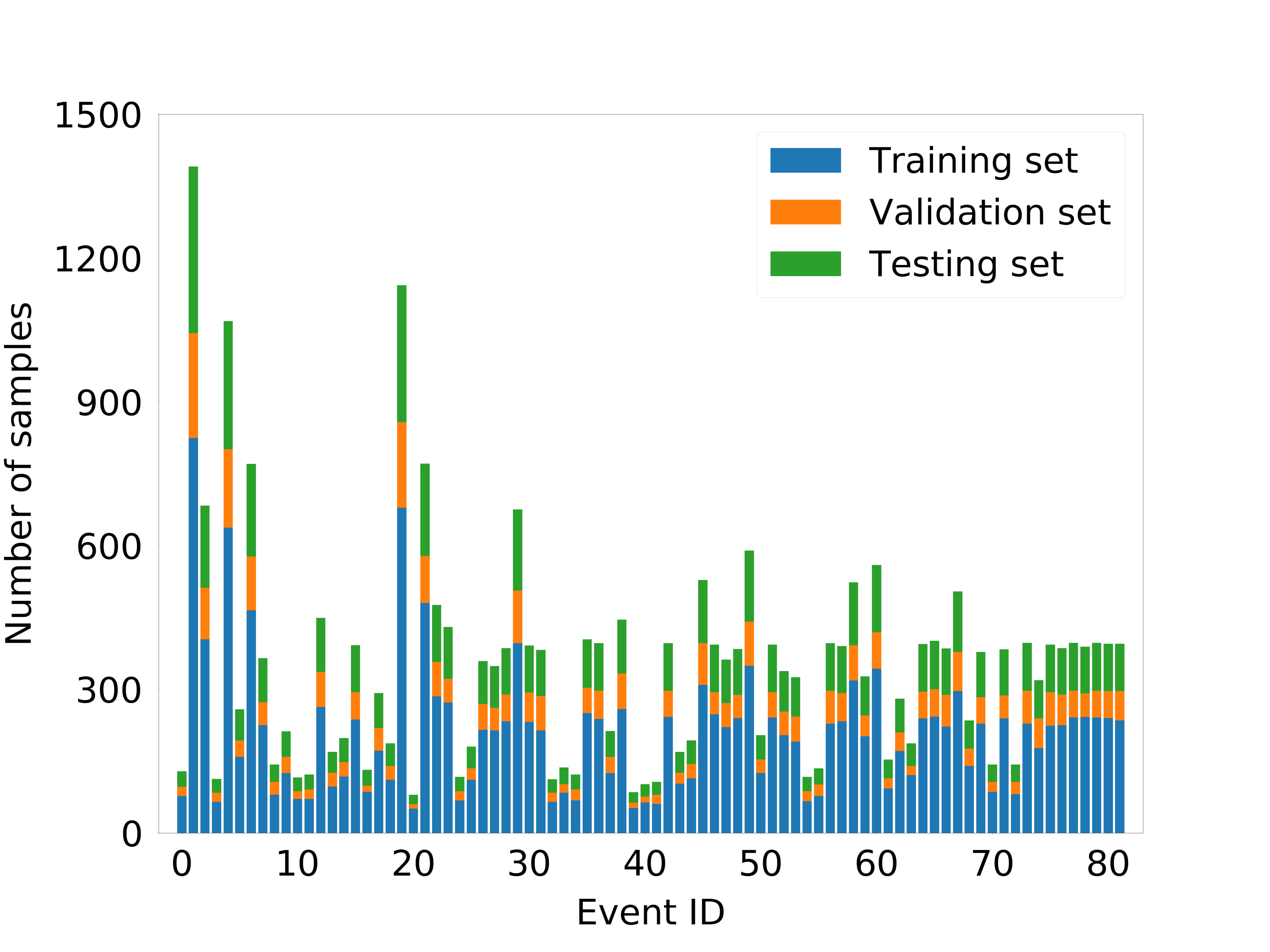}}
\vspace{0pt}
\subfigure[News media articles]{\includegraphics[width=0.48\textwidth]{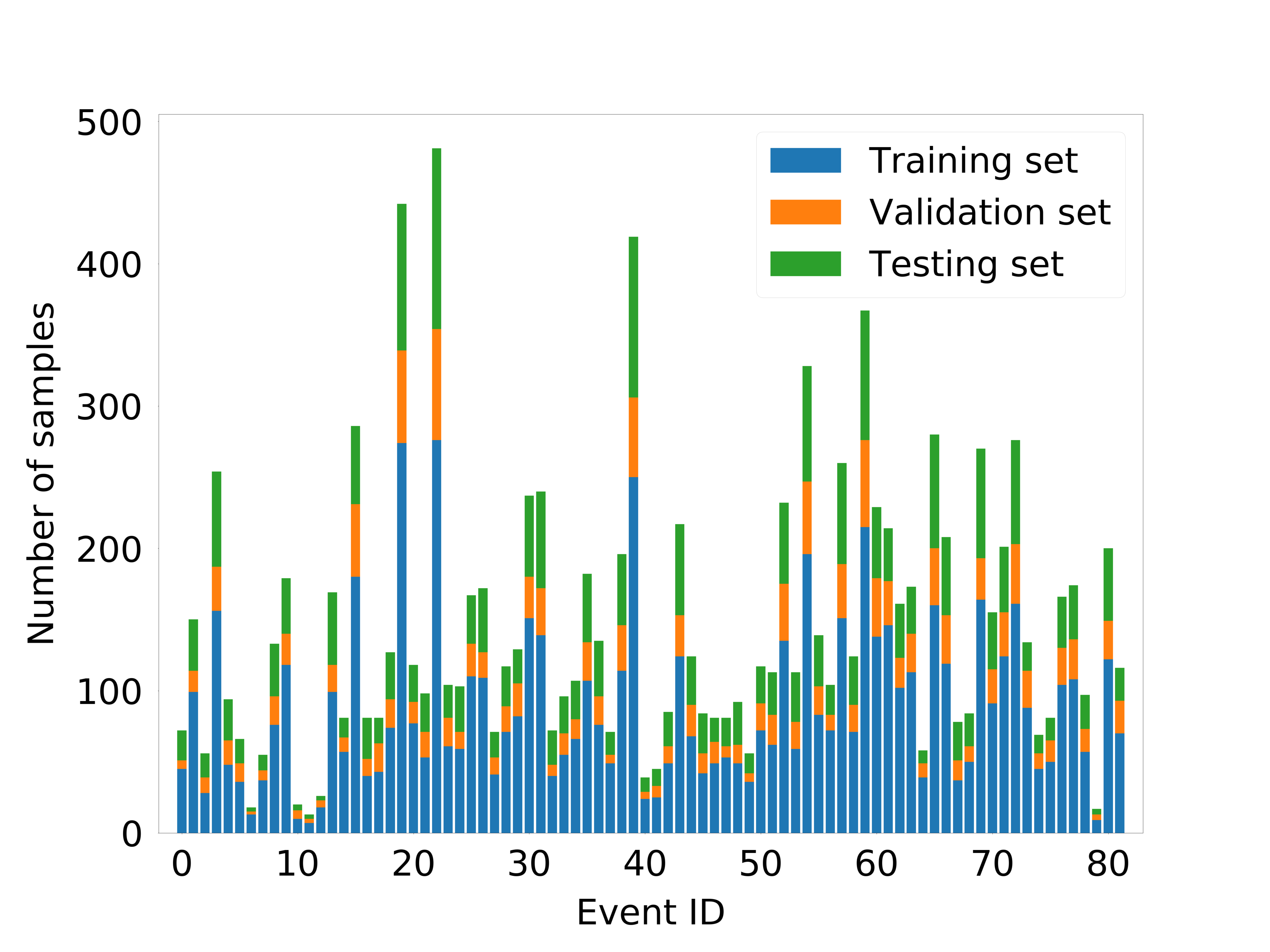}}
\textbf{Fig. 5.} Data distributions of the Wiki-Flickr Event dataset
\end{figure}

\subsection{Evaluation Metric}
In our experiments, we conduct two cross-modal retrieval tasks, i.e., image as query to retrieval texts, and vice versa, which are denoted as Image{$\to$}Text and Text{$\to$}Image, respectively. By following \cite{3}, we evaluate the ranking list by mean average precision (MAP), which is the mean value of average precision (AP) scores of all queries, and {$AP$} is computed as:

\begin{equation}
\label{eqn:L16}
AP = \frac{1}{R}\sum_{k=1}^{n}{\frac{R_k}{k} \times {rel}_k}.
\end{equation}

\noindent where {$R$} denotes relevant item number in test set, {$R_k$} denotes the number of relevant items in top-{$k$} results, and {${rel}_k$} means whether the {$k$}-th result is relevant.

\subsection{Baselines}
For comparisons on paired datasets, i.e., Wikipedia dataset and Pascal Sentence dataset, we include 10 state-of-the-art methods, such as CCA \cite{35}, CFA \cite{6}, KCCA \cite{36}, Corr-AE \cite{18}, DCCA \cite{37}, CMDN \cite{19}, Deep-SM \cite{3}, ACMR \cite{23}, CCL \cite{20}, and ${\rm DSS\ }$\cite{1}. The last seven approaches are representative deep learning models for cross-modal retrieval. All these approaches have reported their performance on these datasets, thus we summarize the best results in these papers for fair comparisons.
For comparisons on unpaired dataset, i.e., Wiki-Flickr Event dataset. We implement a number of baselines for comparisons, including CCA \cite{35}, Deep-SM \cite{3}, ACMR \cite{23}, and DSS \cite{1}. In terms of the implementations, the network structures being adopted may have an influence on the performance. For fairness, we utilize the same VGG19 to extract the 4,096-{$d$} features as the image features, and 3,000 dimensional bag-of-words (BoW) vectors as text features. Distance metric has an influence on the performance. To be consistent with the baselines, we use cosine distance as the distance metric for matching.

\subsection{Implementation Details}
In the experiments, we use two-layer fully-connected networks to project the visual and textual features nonlinearly into a common subspace, i.e., {$R_I$}{$\to$}1000{$\to$}100 for an image and {$R_T$}{$\to$}1000{$\to$}100 for a text. During training, we crop the images and horizontally flip the images randomly with a given probability of 0.5 for data augmentation. The images are resized to 224{$\times$}224 and normalized with mean and standard deviation. The batch size is set as 64, and we use SGD with the momentum as 0.9 and learning rate as 0.01 to optimize parameters.

\subsection{Performance on Cross-modal (Event) Retrieval}

\subsubsection{Cross-modal Retrieval}
Table 3 summarizes the best performance of the baselines reported in their papers on the Wikipedia and Pascal Sentence datasets. From the table, we can draw the following observations. Firstly, the approaches exploiting deep learning models, such as CMDN, Deep-SM, ACMR, CCL, DSS outperform the classical correlation-based models, such as CCA, CFA, KCCA, etc. The experimental results demonstrate the effectiveness of the deep learning models for cross-modal retrieval, benefiting from their abilities of learning and extracting discriminative features. Secondly, our ${S}^{3}CA$ achieves significant improvement on the performance compared to both the traditional and the deep learning methods, benefiting from the joint training of the modality-specific neural networks with CORAL alignment. In particular, ACMR is a seminal work using GANs for cross-modal retrieval, which aligns the distributions between visual and textual features by adopting adversarial learning. As mentioned previously, it still needs further study on the relationships between ``indistinguishable by discriminators" and ``obeying the same data distribution" in the context of cross-modal retrieval. In Section 5.6.4, we will further investigate the adversarial loss and CORAL alignment.

\begin{center}
\textbf{Table 3.} Performance of the approaches on the strongly-aligned paired datasets.
\end{center}

\begin{tabular}{|l|l|c|c|c|c|c|c|}
\hline
\multirow{2}{*}{Dataset}&\multirow{2}{*}{Method}&\multicolumn{3}{|c|}{Task}\\
\cline{3-5}
& &Image{$\to$}Text&Text{$\to$}Image&Average\\
\hline
\multirow{11}{*}{Wikipedia}&CCA&0.258&0.250&0.254\\
 &CFA&0.334&0.297&0.316 \\
 &KCCA&0.215&0.214&0.215 \\
 &Corr-AE&0.402&0.395&0.399 \\
 &DCCA&0.440&0.390&0.415 \\
 &CMDN&0.488&0.427&0.458 \\
 &Deep-SM&0.478&0.422&0.450 \\
 &ACMR&0.468&0.412&0.440 \\
 &CCL&0.504&0.457&0.481 \\
 &DSS&0.516&0.461&0.489 \\
 &${S}^{3}CA$&\bfseries{0.551}&\bfseries{0.485}&\bfseries{0.518} \\
\hline

\multirow{11}{*}{Pascal Sentence}&CCA&0.169&0.151&0.160\\
 &CFA&0.351&0.340&0.346 \\
 &KCCA&0.209&0.192&0.201 \\
 &Corr-AE&0.489&0.484&0.487 \\
 &DCCA&0.456&0.462&0.459 \\
 &CMDN&0.544&0.526&0.535 \\
 &Deep-SM&0.560&0.539&0.550 \\
 &ACMR&0.538&0.544&0.541 \\
 &CCL&0.566&0.560&0.563 \\
 &DSS&0.545&0.574&0.560 \\
 &${S}^{3}CA$&\bfseries{0.588}&\bfseries{0.607}&\bfseries{0.598} \\
\hline
\end{tabular}

\subsubsection{Cross-modal Event Retrieval}
Table 4 summarizes the performance of the baselines on our unpaired Wiki-Flickr Event dataset, from which we can conclude some observations. Firstly, the performance of Deep-SM and ACMR on the unpaired dataset drops more or less compared with the performance on the paired datasets in Table 3 horizontally. The phenomenon indicates the difficulty of the cross-modal event retrieval task on weakly-aligned unpaired dataset. Secondly, our DSS and the improved ${S}^{3}CA$ achieve quite robust performance on both paired and unpaired datasets. In terms of the robustness and MAP metric, the proposed ${S}^{3}CA$ outperforms the baselines obviously.

\begin{center}
\textbf{Table 4.} Performance of the approaches on the weakly-aligned unpaired dataset.
\end{center}

\begin{tabular}{|l|l|c|c|c|c|c|c|}
\hline
\multirow{2}{*}{Dataset}&\multirow{2}{*}{Method}&\multicolumn{3}{|c|}{Task}\\
\cline{3-5}
& &Image{$\to$}Text&Text{$\to$}Image&Average\\
\hline
\multirow{5}{*}{Wiki-Flickr Event}&CCA&0.244&0.218&0.231\\
 &Deep-SM&0.373&0.386&0.380 \\
 &ACMR&0.508&0.481&0.495 \\
 &DSS&0.578&0.570&0.574 \\
 &${S}^{3}CA$&\bfseries{0.608}&\bfseries{0.611}&\bfseries{0.610} \\
\hline
\end{tabular}

\subsection{Further Analyses on ${S}^{3}CA$}

\subsubsection{Convergence in practice}
We visualize the objective function of ${S}^{3}CA$ in Equation (13) with the increasing number of interactions in Fig.6. From the figure, we can see that the objective function of ${S}^{3}CA$ converges on both paired and unpaired datasets after training for a few epochs.

\begin{figure*}[h]
\centering
\setcounter{subfigure}{0}

\subfigure[Wiki-Flickr Event dataset]{\includegraphics[width=0.32\textwidth]{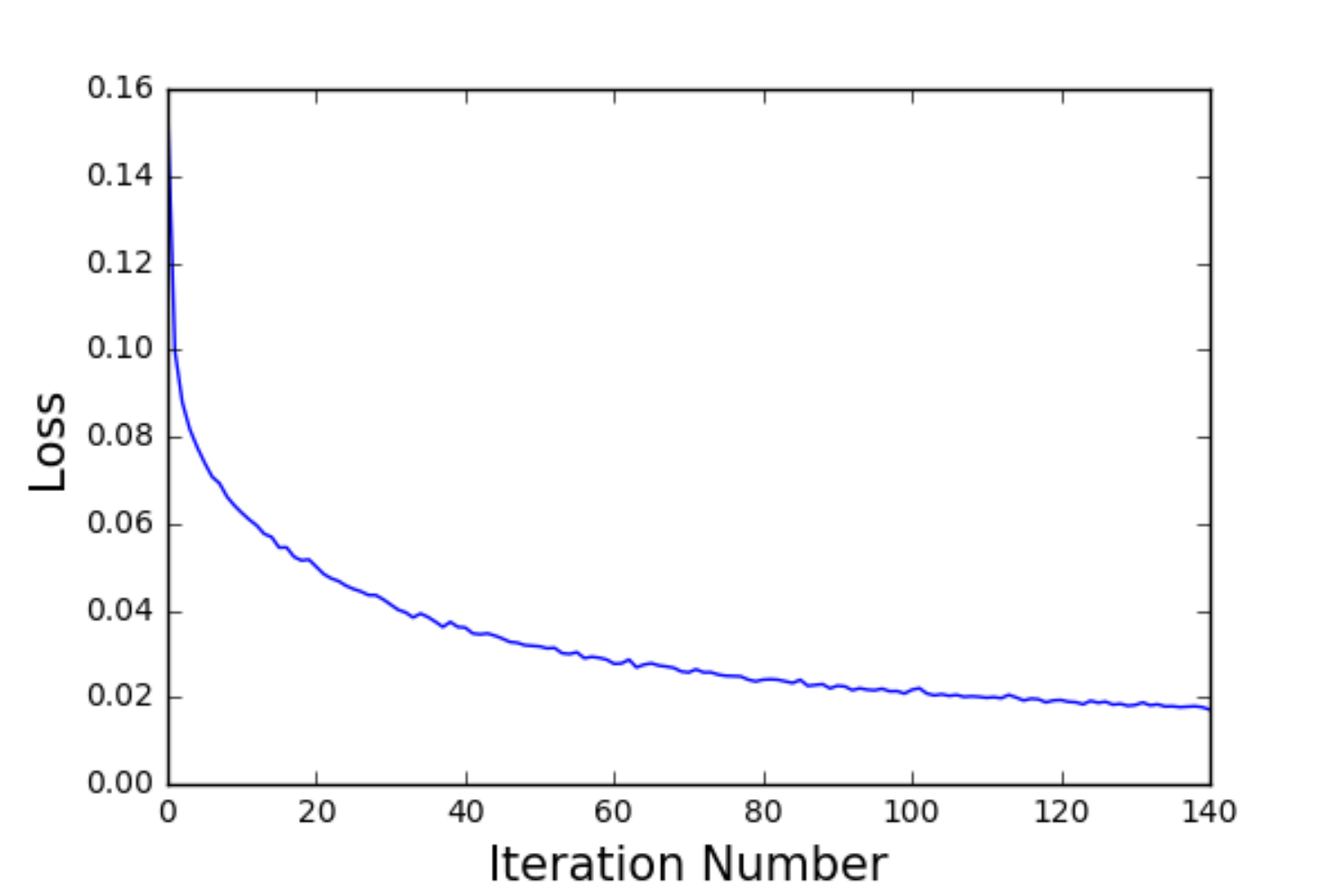}}
\vspace{0pt}
\subfigure[Wikipedia dataset]{\includegraphics[width=0.32\textwidth]{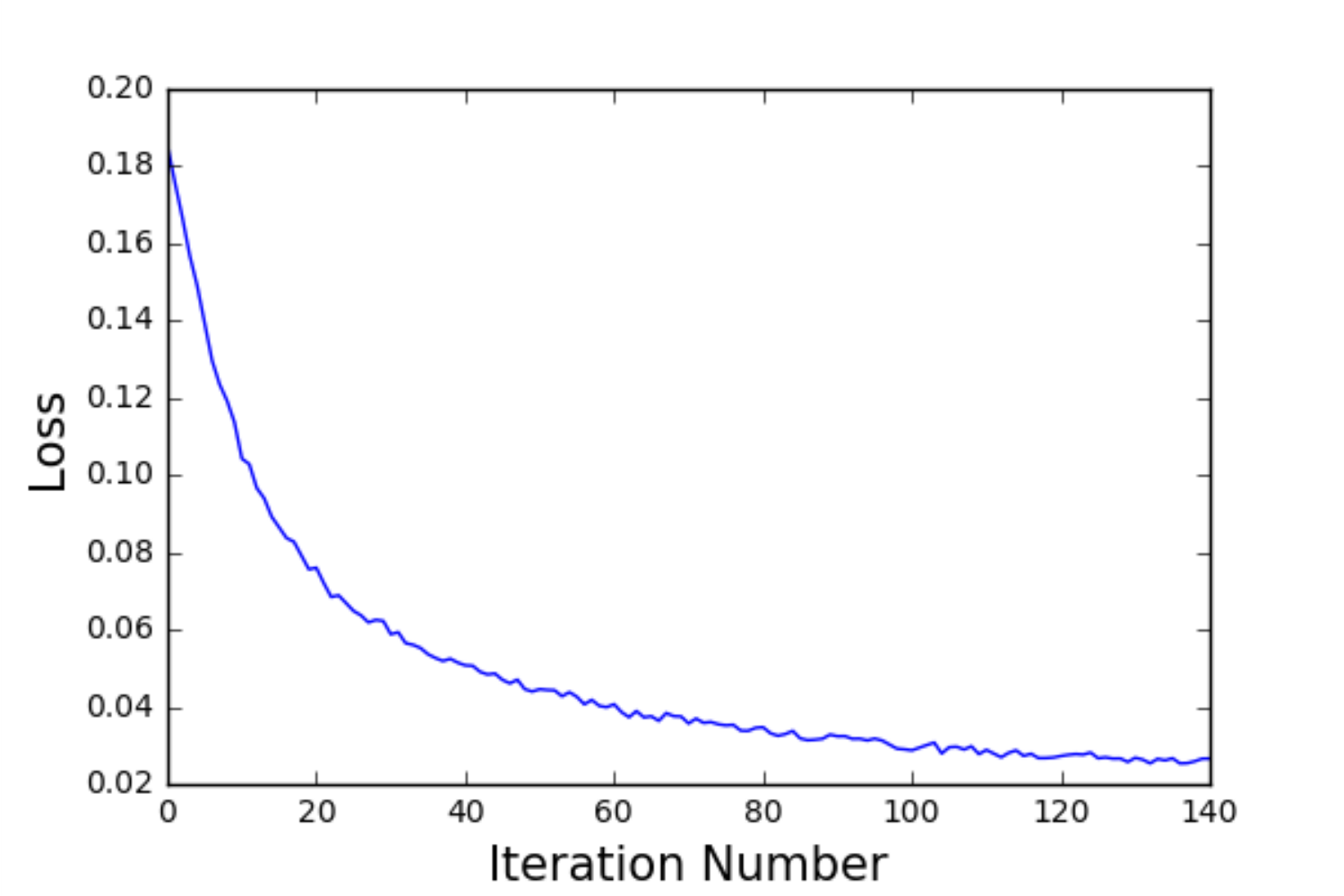}}
\subfigure[Pascal Sentence dataset]{\includegraphics[width=0.32\textwidth]{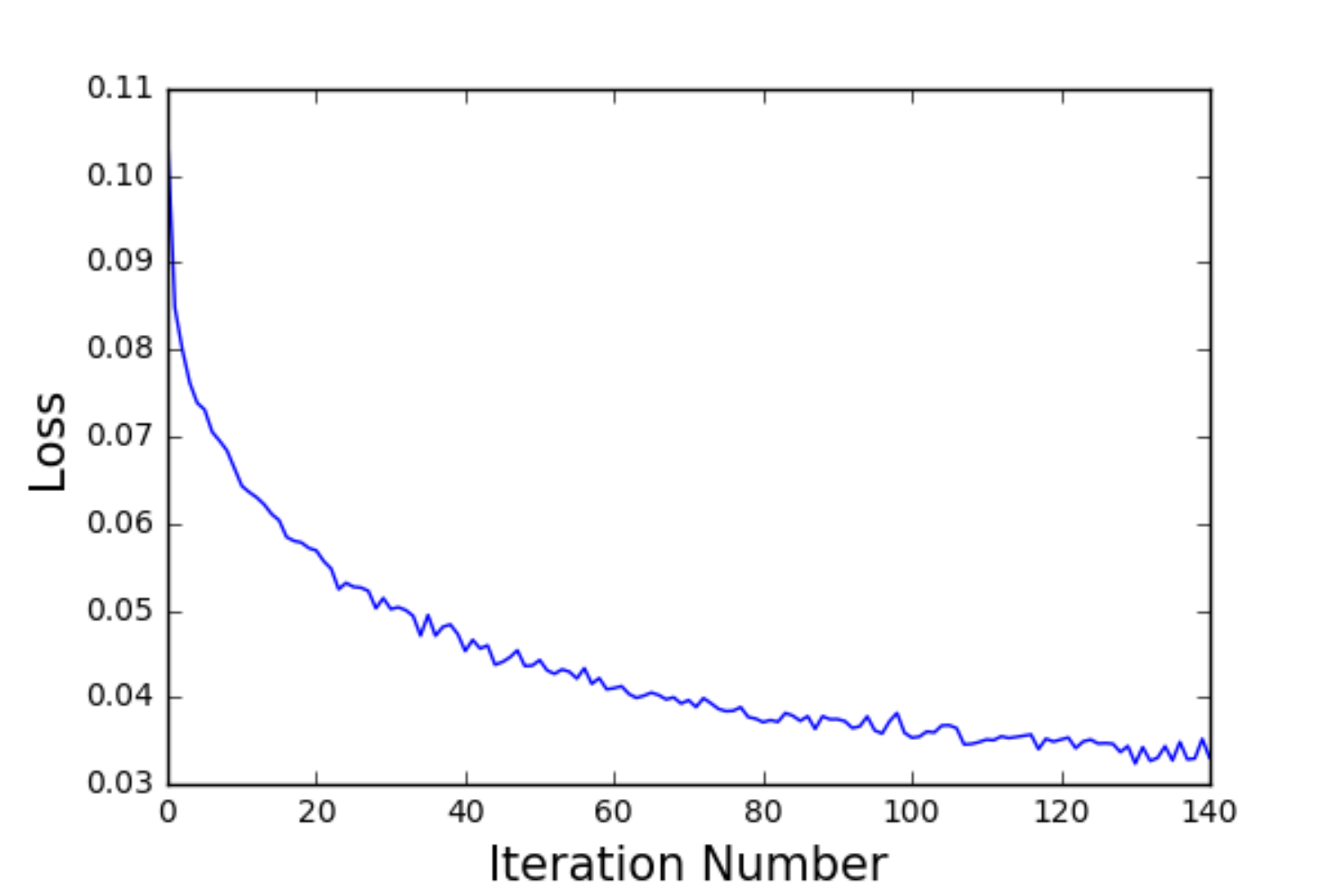}}
\textbf{Fig. 6.} Convergence of ${S}^{3}CA$ in practice
\end{figure*}

\subsubsection{Evaluation on distance metrics}
We evaluate the different metrics that can be used for cross-modal event retrieval. The evaluations are conducted on the Wiki-Flickr Event dataset for illustrations in Table 5, from which we can observe that normalized correlation and cosine distance perform better than Euclidean distance and KL-divergence. For the rest of experiments on the analyses of our ${S}^{3}CA$, we will use normalized correlation as distance metric accordingly.

\begin{center}
\textbf{Table 5.} MAP performance of our ${S}^{3}CA$ adopting different distance metrics on Wiki-Flickr Event dataset.
\end{center}

\begin{center}
\begin{tabular}{|l|c|c|c|}
\hline
\multirow{2}{*}{Distance Metric}&\multicolumn{3}{|c|}{Task}\\
\cline{2-4}
 &Image{$\to$}Text&Text{$\to$}Image&Average\\
\hline
KL-divergence&0.567&0.523&0.545\\
\hline
Euclidean Distance&0.530&0.513&0.522\\
\hline
Cosine Distance&0.608&0.611&0.610\\
\hline
Normalized Correlation&\bfseries{0.625}&\bfseries{0.630}&\bfseries{0.628}\\
\hline
\end{tabular}
\end{center}

\subsubsection{Evaluation on the effectiveness of the CORAL loss}
For illustrations, we take Wiki-Flickr Event dataset as an example to show the CORAL distance when training modality-specific neural networks with or without CORAL, which is shown in Fig. 7-a. We can find that the CORAL distance between the activation features of the modality-specific neural networks increases dramatically after training a few epochs without the CORAL. In contrast, ${S}^{3}CA$ with CORAL achieves more similar layer activations on neural networks trained on texts and images. In addition, we show the MAP performance of ${S}^{3}CA$ without or with CORAL on validation set in Fig. 7-b. From the figure, we can observe that ${S}^{3}CA$ with CORAL tends to achieve better performance. The experimental results manifest the significance of introducing CORAL for shared semantic space learning in the context of cross-modal (event) retrieval.

\begin{figure}[h]
\centering
\setcounter{subfigure}{0}
\subfigure[CORAL distance]{
\includegraphics[width=0.48\textwidth]{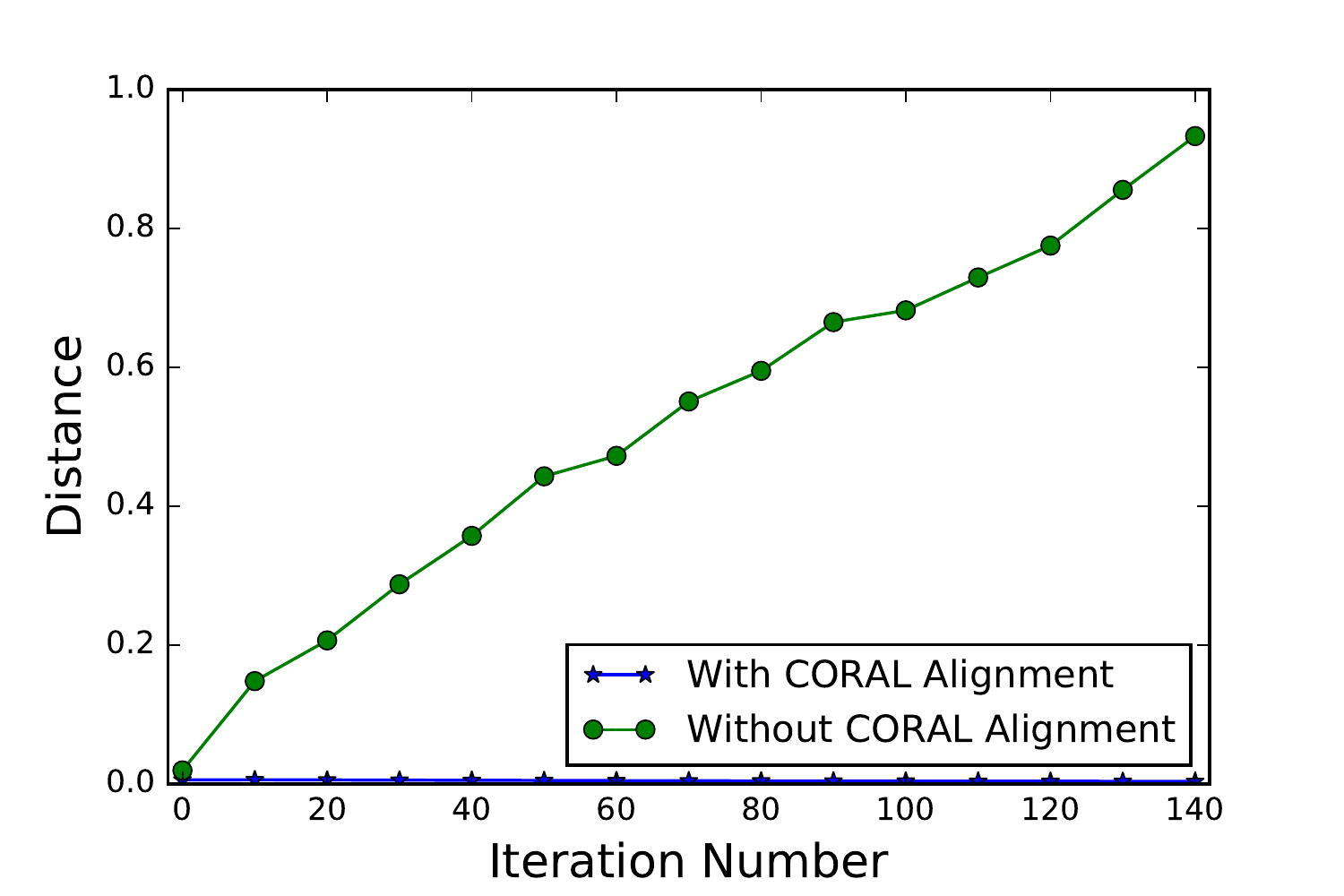}}
\vspace{0pt}
\subfigure[MAP performance]{\includegraphics[width=0.48\textwidth]{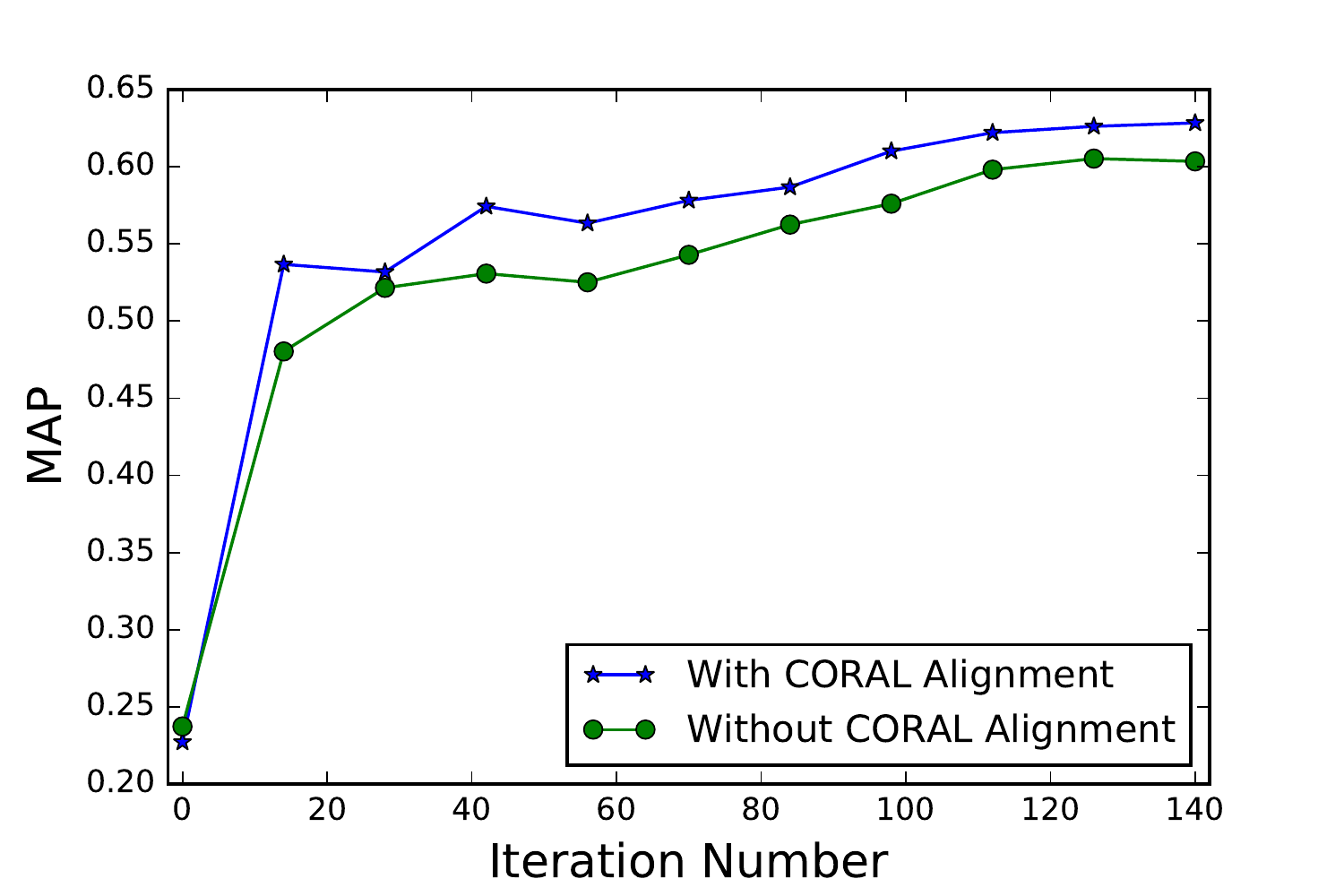}}
\end{figure}

\noindent \textbf{Fig. 7.} Impact of ${S}^{3}CA$ with or without CORAL alignment on CORAL distance and MAP performance (Wiki-Flickr Event dataset).

\subsubsection{Evaluation on the effectiveness of interactive regularization terms}
As mentioned in Section 4.3, there exist a number of interactive regularization terms in addition to CORAL that are available for the joint training of the modality-specific neural networks. Consequently, we implement four variants of our ${S}^{3}CA$ by using four different interactive regularization terms, respectively, including triplet loss, adversarial loss, Maximum Mean Discrepancy (MMD) loss, and CORAL. The experimental results on both paired and unpaired datasets are summarized in Table 6, where ``N.A." denotes no interactive regularization terms being used. From the table, we can conclude two observations. Firstly, most approaches adopting interactive regularization terms achieve improvement on the performance. In particular, the triplet loss not only relies on high-quality data pairs for training, but also needs to select appropriate strategies for choosing positive and negative samples. We have tries a few different strategies on selecting positive and negative data pairs for triplet loss, such as selecting all the positive and negative data pairs in a batch, or selecting the close data pairs, or selecting the data pairs randomly. The experimental results on the Wiki-Flickr Event dataset make not much difference on the performance. Secondly, ${S}^{3}CA$ with correlation alignment (CORAL) achieves the best performance on both paired datasets and unpaired dataset.

\begin{center}
\

\textbf{Table 6.} MAP scores of our ${S}^{3}CA$ combined with interactive regularization terms.
\end{center}

\begin{tabular}{|l|l|c|c|c|c|c|c|}
\hline
\multirow{2}{*}{Dataset}&\multirow{2}{*}{Constraint}&\multicolumn{3}{|c|}{Task}\\
\cline{3-5}
& &Image{$\to$}Text&Text{$\to$}Image&Average\\
\hline
\multirow{5}{*}{Wikipedia}&N.A.&0.539&0.478&0.508\\
\cline{2-5}
&Triplet&0.495&0.479&0.487\\
\cline{2-5}
&Adversarial&0.548&0.487&0.517\\
\cline{2-5}
&MMD&0.543&0.473&0.508\\
\cline{2-5}
&CORAL&\bfseries{0.564}&\bfseries{0.487}&\bfseries{0.526}\\
\hline

\multirow{5}{*}{Pascal Sentence}&N.A.&0.551&0.579&0.565\\
\cline{2-5}
&Triplet&0.535&0.543&0.539\\
\cline{2-5}
&Adversarial&0.569&0.589&0.579\\
\cline{2-5}
&MMD&0.571&0.604&0.587\\
\cline{2-5}
&CORAL&\bfseries{0.596}&\bfseries{0.619}&\bfseries{0.607}\\
\hline

\multirow{5}{*}{Wiki-Flickr Event}&N.A.&0.610&0.608&0.609\\
\cline{2-5}
&Triplet&0.610&0.612&0.611\\
\cline{2-5}
&Adversarial&0.613&0.606&0.609\\
\cline{2-5}
&MMD&0.610&0.615&0.612\\
\cline{2-5}
&CORAL&\bfseries{0.625}&\bfseries{0.630}&\bfseries{0.628}\\
\hline
\end{tabular}

\subsubsection{Evaluation on new events in test set}
 Multimodal data fusion is a critical issue, beyond which real-world events are expected to be retrieved crossing different data domains. In reality, a special case may be that the query from a test set is about a new event that has never appears in the training and validation sets, which can be called new event retrieval. Therefore, we evaluate ${S}^{3}CA$ by removing all the data samples related to some events from the training and validation sets, while the samples with these events labels in the test set are still used as queries, i.e., there are new events in the test set. The performance of ${S}^{3}CA$ on new event retrieval is shown in Fig. 8. From the figure, we can observe that unknown event labels in the testing set may decrease the performance of ${S}^{3}CA$, which is still need to explore and improve.

\begin{figure}[h]
\centering
\includegraphics[width=0.7\textwidth]{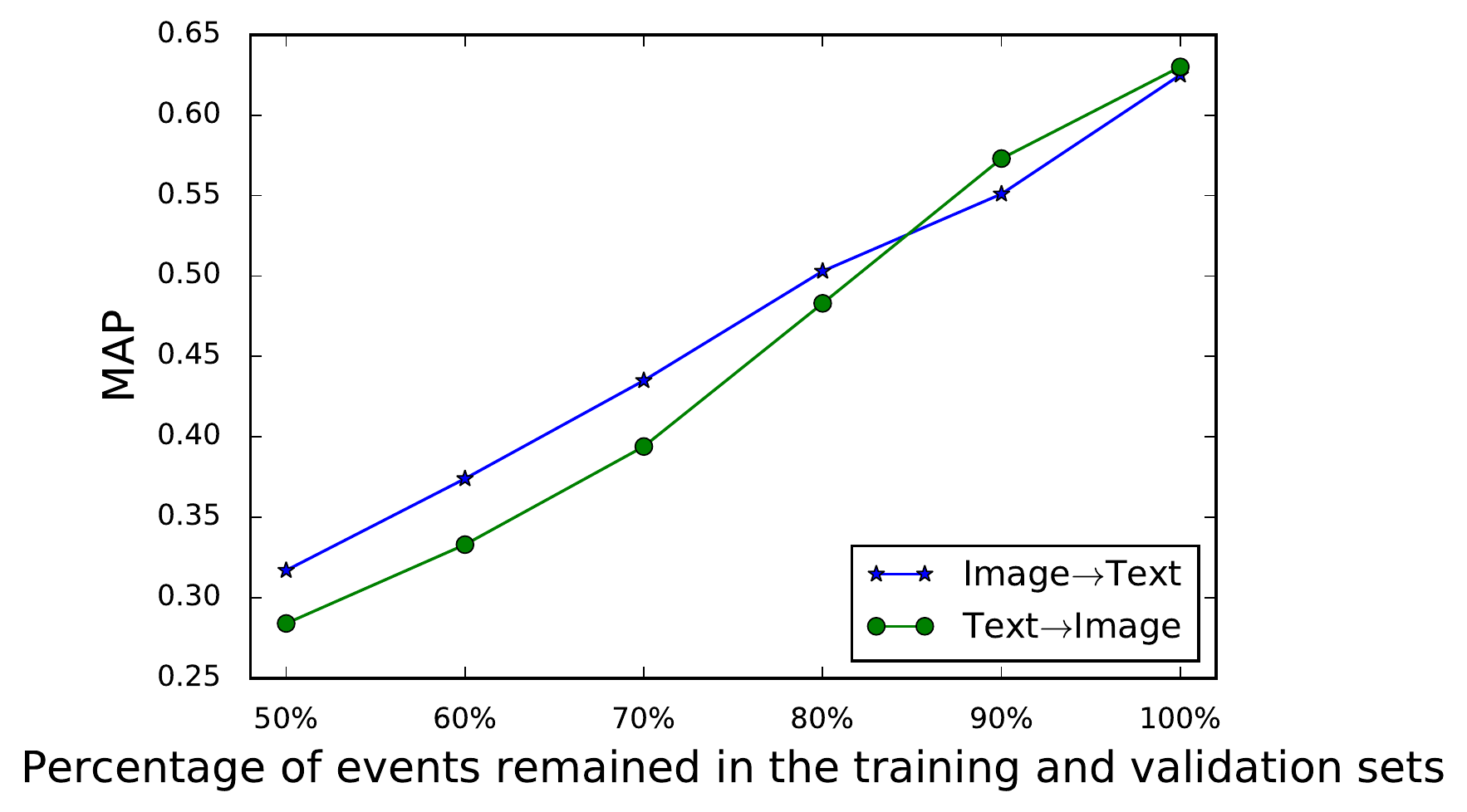}
\end{figure}

\noindent \textbf{Fig. 8.} Performance of ${S}^{3}CA$ on cross-modal new event retrieval.

\subsection{Examples of the Retrieved Results by ${S}^{3}CA$}
Intuitively, we take text retrieving images as an example to show the performance of ${S}^{3}CA$ on Wiki-Flickr Event dataset in Fig. 9. The top-5 images are given in the figure, where the event labels are marked at the lower right corner. Red boxes indicate the mismatched retrieved results, while green boxes indicate the correct results. Our ${S}^{3}CA$ returns one mismatched image in the third example and three mismatched images in the last example. In the third example, both events of ``2013 Savar building collapse" and ``Shooting of Michael Brown" have injured people, sharing images that are quite similar visually, which are difficult to distinguish. In the last example, the event of ``2014 Hong Kong protests", ``Umbrella Movement" and ``Sunflower Student Movement" have some overlap on the date. Though the three events have different entries in Wikipedia indicating that they are different event labels, yet they almost share the same content in terms of event elements, such as when, where, who, what, how, why.

\begin{center}
\includegraphics[width=\textwidth]{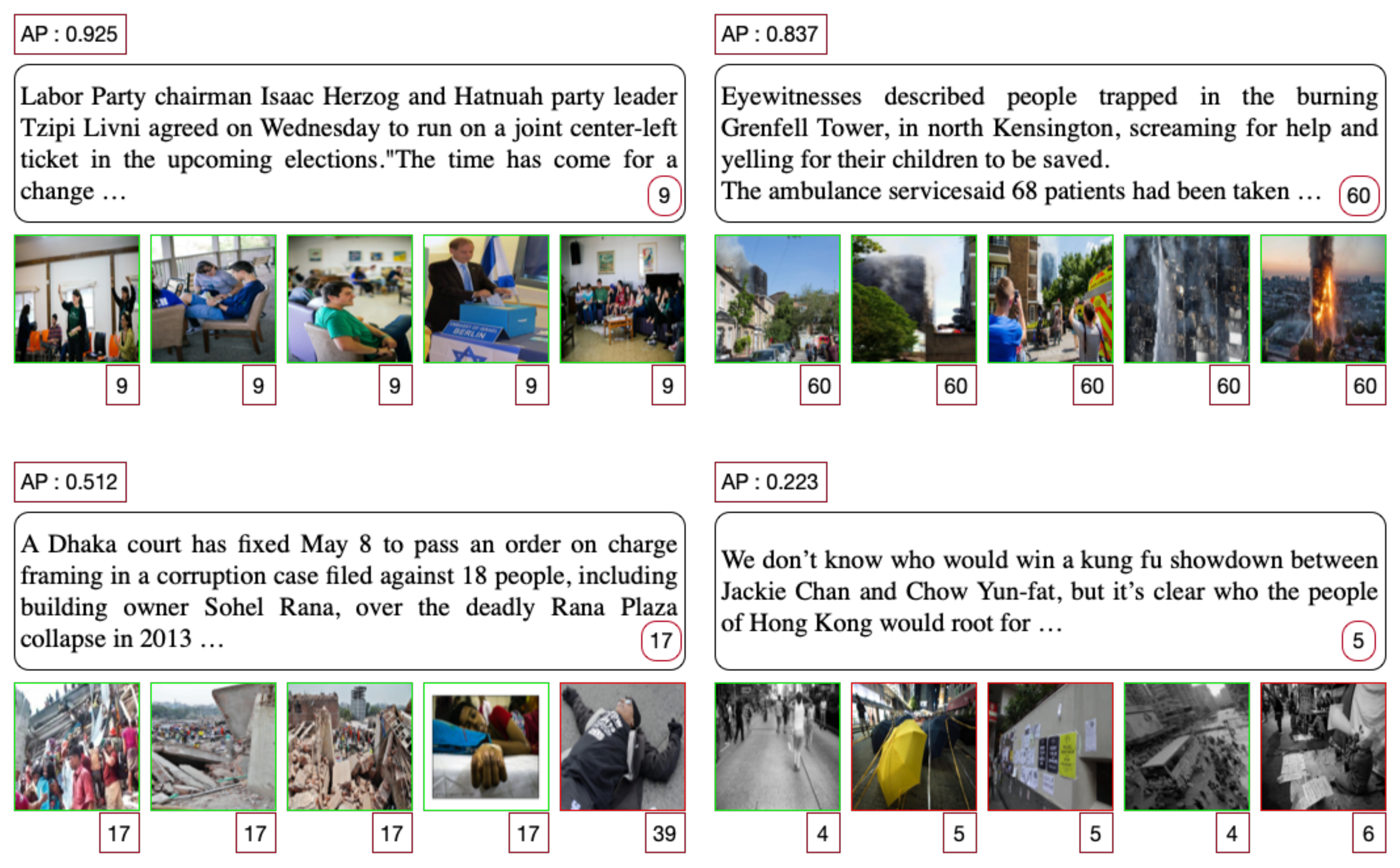}
\end{center}

\noindent \textbf{Fig. 9.} Four examples of cross-modal retrieval results obtained by ${S}^{3}CA$ on Wiki-Flickr Event dataset. Note that the numbers refer to the event labels (i.e., 4: ``2014 Hong Kong protests", 5: ``Umbrella Movement", 6: ``Sunflower Student Movement", 9: ``Israeli legislative election, 2015", 17: ``2013 Savar building collapse", 39: ``Shooting of Michael Brow", 60: ``Grenfell Tower fire", etc.)

\section{Conclusions}
In this paper, we have proposed a model of learning shared semantic space with correlation alignment (${S}^{3}CA$) for cross-modal (event) retrieval. ${S}^{3}CA$ embeds multimodal data into a shared semantic space with high-level semantics, and utilizes the correlation alignment (CORAL) to align the distributions between the layer activations in the two neural networks trained on images and texts. We contribute a weakly-aligned unpaired Wiki-Flickr Event dataset as a complement of the existing paired datasets for cross-modal retrieval. To the best of our knowledge, it is the first work investigating cross-modal retrieval tasks on unpaired data, especially focusing on real-world events. Extensive experiments conducted on both paired datasets and unpaired Wiki-Flickr Event dataset show the superiority of ${S}^{3}CA$, outperforming the state-of-the-art approaches.

\section*{References}

\bibliography{mybibfile}

\begin{thebibliography}{10}
\expandafter\ifx\csname url\endcsname\relax
  \def\url#1{\texttt{#1}}\fi
\expandafter\ifx\csname urlprefix\endcsname\relax\def\urlprefix{URL }\fi
\expandafter\ifx\csname href\endcsname\relax
  \def\href#1#2{#2} \def\path#1{#1}\fi

\bibitem{1}
R.~Situ, Z.~Yang, J.~Lv, Q.~Li, W.~Liu, Cross-modal event retrieval: A dataset
  and a baseline using deep semantic learning, in: Pacific Rim Conference on
  Multimedia, Springer, 2018, pp. 147--157.

\bibitem{2}
Q.-Y. Jiang, W.-J. Li, Deep cross-modal hashing, 2016.

\bibitem{3}
Y.~Wei, Y.~Zhao, C.~Lu, S.~Wei, L.~Liu, Z.~Zhu, S.~Yan, Cross-modal retrieval
  with cnn visual features: A new baseline, Vol.~47, IEEE, 2017, pp. 449--460.

\bibitem{4}
C.~Li, C.~Deng, N.~Li, W.~Liu, X.~Gao, D.~Tao, Self-supervised adversarial
  hashing networks for cross-modal retrieval, in: Proceedings of the IEEE
  Conference on Computer Vision and Pattern Recognition, 2018, pp. 4242--4251.

\bibitem{5}
D.~R. Hardoon, S.~Szedmak, J.~Shawe-Taylor, Canonical correlation analysis: An
  overview with application to learning methods, Vol.~16, MIT Press, 2004, pp.
  2639--2664.

\bibitem{6}
D.~Li, N.~Dimitrova, M.~Li, I.~K. Sethi, Multimedia content processing through
  cross-modal association, in: Proceedings of the eleventh ACM international
  conference on Multimedia, ACM, 2003, pp. 604--611.

\bibitem{7}
N.~Rasiwasia, J.~Costa~Pereira, E.~Coviello, G.~Doyle, G.~R. Lanckriet,
  R.~Levy, N.~Vasconcelos, A new approach to cross-modal multimedia retrieval,
  in: Proceedings of the 18th ACM international conference on Multimedia, ACM,
  2010, pp. 251--260.

\bibitem{8}
A.~Sharma, A.~Kumar, H.~Daume, D.~W. Jacobs, Generalized multiview analysis: A
  discriminative latent space, in: Computer Vision and Pattern Recognition
  (CVPR), 2012 IEEE Conference on, IEEE, 2012, pp. 2160--2167.

\bibitem{9}
Y.~Gong, Q.~Ke, M.~Isard, S.~Lazebnik, A multi-view embedding space for
  modeling internet images, tags, and their semantics, Vol. 106, Springer,
  2014, pp. 210--233.

\bibitem{10}
V.~Ranjan, N.~Rasiwasia, C.~Jawahar, Multi-label cross-modal retrieval, in:
  Proceedings of the IEEE International Conference on Computer Vision, 2015,
  pp. 4094--4102.

\bibitem{11}
T.~Quynh Nhi~Tran, H.~Le~Borgne, M.~Crucianu, Aggregating image and text
  quantized correlated components, in: Proceedings of the IEEE Conference on
  Computer Vision and Pattern Recognition, 2016, pp. 2046--2054.

\bibitem{12}
B.~Bai, J.~Weston, D.~Grangier, R.~Collobert, K.~Sadamasa, Y.~Qi, O.~Chapelle,
  K.~Weinberger, Learning to rank with (a lot of) word features, Vol.~13,
  Springer, 2010, pp. 291--314.

\bibitem{13}
L.~Wang, Y.~Li, S.~Lazebnik, Learning deep structure-preserving image-text
  embeddings, in: Proceedings of the IEEE conference on computer vision and
  pattern recognition, 2016, pp. 5005--5013.

\bibitem{14}
L.~Wang, Y.~Li, J.~Huang, S.~Lazebnik, Learning two-branch neural networks for
  image-text matching tasks, IEEE, 2018.

\bibitem{15}
A.~Salvador, N.~Hynes, Y.~Aytar, J.~Marin, F.~Ofli, I.~Weber, A.~Torralba,
  Learning cross-modal embeddings for cooking recipes and food images, Vol.
  720, 2017, p.~2.

\bibitem{16}
L.~Zhang, B.~Ma, G.~Li, Q.~Huang, Q.~Tian, Multi-networks joint learning for
  large-scale cross-modal retrieval, in: Proceedings of the 2017 ACM on
  Multimedia Conference, ACM, 2017, pp. 907--915.

\bibitem{17}
F.~Yan, K.~Mikolajczyk, Deep correlation for matching images and text, in:
  Proceedings of the IEEE conference on computer vision and pattern
  recognition, 2015, pp. 3441--3450.

\bibitem{18}
F.~Feng, X.~Wang, R.~Li, Cross-modal retrieval with correspondence autoencoder,
  in: Proceedings of the 22nd ACM international conference on Multimedia, ACM,
  2014, pp. 7--16.

\bibitem{19}
Y.~Peng, X.~Huang, J.~Qi, Cross-media shared representation by hierarchical
  learning with multiple deep networks., in: IJCAI, 2016, pp. 3846--3853.

\bibitem{20}
Y.~Peng, J.~Qi, X.~Huang, Y.~Yuan, Ccl: Cross-modal correlation learning with
  multigrained fusion by hierarchical network, Vol.~20, IEEE, 2018, pp.
  405--420.

\bibitem{21}
Y.~He, S.~Xiang, C.~Kang, J.~Wang, C.~Pan, Cross-modal retrieval via deep and
  bidirectional representation learning, Vol.~18, IEEE, 2016, pp. 1363--1377.

\bibitem{22}
M.~Fan, W.~Wang, P.~Dong, L.~Han, R.~Wang, G.~Li, Cross-media retrieval by
  learning rich semantic embeddings of multimedia, in: Proceedings of the 2017
  ACM on Multimedia Conference, ACM, 2017, pp. 1698--1706.

\bibitem{23}
B.~Wang, Y.~Yang, X.~Xu, A.~Hanjalic, H.~T. Shen, Adversarial cross-modal
  retrieval, in: Proceedings of the 2017 ACM on Multimedia Conference, ACM,
  2017, pp. 154--162.

\bibitem{24}
X.~Zhang, S.~Zhou, J.~Feng, H.~Lai, B.~Li, Y.~Pan, J.~Yin, S.~Yan, Hashgan:
  Attention-aware deep adversarial hashing for cross modal retrieval, 2017.

\bibitem{25}
K.~Simonyan, A.~Zisserman, Very deep convolutional networks for large-scale
  image recognition, 2014.

\bibitem{26}
B.~Sun, K.~Saenko, Deep coral: Correlation alignment for deep domain
  adaptation, in: European Conference on Computer Vision, Springer, 2016, pp.
  443--450.

\bibitem{27}
B.~Sun, J.~Feng, K.~Saenko, Return of frustratingly easy domain adaptation.,
  in: AAAI, Vol.~6, 2016, p.~8.

\bibitem{28}
F.~Faghri, D.~J. Fleet, J.~R. Kiros, S.~Fidler, Vse++: Improved visual-semantic
  embeddings, Vol.~2, 2017, p.~8.

\bibitem{29}
K.~M. Borgwardt, A.~Gretton, M.~J. Rasch, H.-P. Kriegel, B.~Sch{\"o}lkopf,
  A.~J. Smola, Integrating structured biological data by kernel maximum mean
  discrepancy, Vol.~22, Oxford University Press, 2006, pp. e49--e57.

\bibitem{30}
I.~Goodfellow, J.~Pouget-Abadie, M.~Mirza, B.~Xu, D.~Warde-Farley, S.~Ozair,
  A.~Courville, Y.~Bengio, Generative adversarial nets, in: Advances in neural
  information processing systems, 2014, pp. 2672--2680.

\bibitem{31}
T.~Che, Y.~Li, A.~P. Jacob, Y.~Bengio, W.~Li, Mode regularized generative
  adversarial networks, 2016.

\bibitem{32}
S.~Arora, R.~Ge, Y.~Liang, T.~Ma, Y.~Zhang, Generalization and equilibrium in
  generative adversarial nets (gans), 2017.

\bibitem{33}
S.~Arora, Y.~Zhang, Do gans actually learn the distribution? an empirical
  study, 2017.

\bibitem{Zhenguo2019}
Z.~Yang, Q.~Li, W.~Liu, J.~Lv, Shared multi-view data representation for
  multi-domain event detection, in: IEEE Transactions on Pattern Analysis and
  Machine Intelligence (TPAMI), 2019 (Accepted).

\bibitem{34}
C.~Rashtchian, P.~Young, M.~Hodosh, J.~Hockenmaier, Collecting image
  annotations using amazon's mechanical turk, in: Proceedings of the NAACL HLT
  2010 Workshop on Creating Speech and Language Data with Amazon's Mechanical
  Turk, Association for Computational Linguistics, 2010, pp. 139--147.

\bibitem{35}
H.~Hotelling, Relations between two sets of variates, Vol.~28, JSTOR, 1936, pp.
  321--377.

\bibitem{36}
D.~R. Hardoon, S.~Szedmak, J.~Shawe-Taylor, Canonical correlation analysis: An
  overview with application to learning methods, Vol.~16, MIT Press, 2004, pp.
  2639--2664.

\bibitem{37}
F.~Yan, K.~Mikolajczyk, Deep correlation for matching images and text, in:
  Proceedings of the IEEE conference on computer vision and pattern
  recognition, 2015, pp. 3441--3450.

\end{thebibliography}

\end{document}